\documentclass[10pt]{iopart}
\newcommand{\nid}{\noindent}

\newcommand{\beq}{\begin{equation}}
\newcommand{\eeq}{\end{equation}}
\newcommand{\bea}{\begin{eqnarray}}
\newcommand{\eea}{\end{eqnarray}}
\newcommand{\ds}{\displaystyle}
\usepackage{graphicx}
\usepackage{cite}
\usepackage{epsfig}
\usepackage{here}
\usepackage{color}
\usepackage{caption}
\usepackage[usenames,dvipsnames]{pstricks}
\usepackage{amsfonts}
\usepackage{pst-grad} 
\usepackage{pst-plot} 

\begin{document}

\title[Fourth order integration method]{Fourth order indirect integration method for black hole perturbations: even modes}

\author{Patxi Ritter$^{1,2}$ Alessandro D.A.M. Spallicci$^{1}$\footnote{Corresponding author: spallicci@cnrs-orleans.fr}\\
Sofiane Aoudia$^{3}$ St\'ephane Cordier$^{2}$}

\address{$^{1}$ Universit\'e d'Orl\'eans, Observatoire des Sciences de l'Univers en r\'egion Centre, \\ 
LPC2E Campus CNRS, 3A Av. Recherche Scientifique, 45071 Orl\'eans, France}

\address{$^{2}$ Universit\'e d'Orl\'eans, Laboratoire de Math\'ematiques - Analyse, Probabilit\'es, Mod\`elisation - 
Orl\'eans, MAPMO, Rue de Chartres, 45067 Orl\'eans, France}

\address{$^{3}$ Max Planck Institut f\"ur Gravitationphysik, A. Einstein, Am M\"uhlenberg 1, 14476 Potsdam, Deutschland}


\begin{abstract}
On the basis of a recently proposed strategy of finite element integration in time domain for partial differential equations with a singular source term, we present a fourth order 
algorithm for non-rotating black hole perturbations in the Regge-Wheeler gauge. Herein, we address even perturbations induced  by a particle plunging in.   
The forward time value at the upper node of the { $(r^*,t$)} grid cell is obtained by an algebraic sum of i) the preceding node values of the same cell, ii) analytic expressions, related to the jump conditions on the wave function and its derivatives, iii) the values of the wave function at adjacent cells. In this approach, the numerical integration does not deal with the source and potential terms directly, for cells crossed by the particle world line. This scheme has also been applied to circular and eccentric orbits and it will be object of a forthcoming publication.  
\end{abstract}

\date{17 April 2011}

\section{Introduction}

In the scenario of the capture of compact objects by a supermassive black hole of mass $M$, the seized object is compared to a small mass $m$ (henceforth the particle or the source) perturbing the background spacetime curvature and generating gravitational radiation.  
A comprehensive introduction to the general relativistic issues related to EMRI (Extreme Mass Ratio Inspiral) sources is contained in a topical volume \cite{blspwh11}.

Schwarzschild-Droste (SD) \cite{dr15, dr16a, dr16b, sc16} (see Rothman \cite{ro02} for a justification of this terminology), black hole perturbations have been hugely developed in the Regge-Wheeler (RW) gauge, before in vacuum \cite{rewh57} and after in the presence of a particle by Zerilli \cite{ze69, ze70a, ze70b, ze70c}.
The first finite difference scheme in time domain was proposed by Lousto and Price \cite{lopr97b}. The initial conditions, reflecting the past motion of the particle and the initial amount of gravitational waves, were parametrised by Martel and Poisson \cite{mapo02}.  

If the gravitational radiation emitted and the mass of the captured object are to be taken into account for the determination of the  motion of the latter, it is necessary to compute the derivatives of the perturbations { that implies the third derivative of the wave function $\Psi(r^*,t)$, see e.g. {\cite{sp11}.} For a given accuracy ${\cal O}(h)$ of the third derivative of $\Psi$, the error on $\Psi$ itself should be ${\cal O}(h^4)$. Effectively, the reminder ought to be ${\cal O}(h^5)$ due to the presence in the mesh of the particle that lowers by one more degree the convergence order of the code for geometrical effects \cite{ha07}. We have therefore developed a fourth order scheme.}

The complexity in assessing the continuity of the perturbations at the position of the particle and the compatibility of the self-force to the harmonic (Lorenz-de Donder\footnote{FitzGerald is considered to have also identified the harmonic gauge \cite{hu91}.}) gauge \cite{lo67, dd21} has led researchers to convey their efforts to this gauge, as commenced by Barack and Lousto \cite{balo05}. Conversely, work in harmonic gauge is made cumbersome by the presence of a system of ten coupled equations which replace the single wave equation of the RW gauge. 

We have proposed \cite{aosp11a, aosp11b} a finite element method of integration{ , in RW gauge,} based on the jump conditions that the wave function and its derivatives have to satisfy for the SD black hole perturbations to be continuous at the position of the particle. We first deal with the radial trajectory and the associated even parity perturbations, while in a forthcoming paper we shall present the circular and eccentric orbital cases, referring thus to both odd and even parity perturbations. 

The main feature of this method consists in avoiding the direct and explicit integration { of the wave equation (the potential and the source term with the associated singularities)} whenever the grid cells are crossed by the particle. Indeed, the information on the wave equation is implicitly given by the jump conditions on the wave function and its derivatives. Conversely, for cells not crossed by the particle world line, the integrating method might retain the previous approach by Lousto \cite{lo05} {  and Haas \cite{ha07}. Among the efforts using jump discontinuities, although in a different context, it is worthwhile to mention those of Haas \cite{ha07}, Sopuerta and coworkers \cite{sola06, caso09, caso11} getting the  self-force in a scalar case. For the geodesic gravitational case, 
like Sopuerta and coworkers, Jung et al. \cite{jukhna07}, Chakraborty et al. \cite{chjukh11} rely on spectral methods; 
Zumbusch \cite{zu09}, Field et al. \cite{fihela09} use a discontinuous Galerkin method; Hopper and Evans \cite{hoev10} work partially in frequency domain. Among recent results not based on jump discontinuities but concerning fourth order time domain codes, the one proposed by Thornburg \cite{th10} deals with and adaptive mesh refinement, while Nagar and coworkers replace the delta distribution with a narrow Gaussian \cite{nadata07, bena10}}.  

For the computation of the back-action, this method ensures a well behaved wave function at the particle position, since the approach is governed by the analytical values of the jump conditions at the particle position. 

In \cite{aosp11b} we have provided waveforms at infinity and the wave function at the position of the particle at first order. 
Herein, we focus instead on the improvement of { the} algorithm at fourth order and refer to \cite{aosp11b} for all complementary information. 
The features of this method can be summarised as follows:

\begin{itemize}
\item {Avoidance of direct and explicit integration of the wave equation { (the potential and the source term with the associated singularities)} for the grid cells crossed by the particle.}
\item {Improvement of the reliability, since analytic expressions partly replace numerical ones (the replacement is total at first order \cite{aosp11a,aosp11b}{ )}.}
\item {Applicability of the method to generic orbits, assuming that the even and odd wave equations 
are satisfied by $\Psi$, respectively $R$, being $C^{-1}$\footnote{A $C^{-1}$ continuity class element, like a Heaviside step distribution, may be seen as an element which after integration transforms into an element belonging to the $C^{0}$ class of functions.}.}
\end{itemize}

Geometric units ($G = c = 1$) are used, unless stated otherwise. The metric signature is $(-, +, +, +)$.

\section{The wave equation}

The wave function (its dimension is such that the energy is proportional to $\int_0^\infty
  \dot\Psi^2\,dt$), in the Moncrief form \cite{mo74} and RW gauge \cite{rewh57}, is defined by 

\beq
\Psi_l (t,r)= \frac {r}{\lambda +1}
\left[ K^l+\frac{r-2M}{\lambda r+3M}\left(
H_2^l-r\frac{\partial K^l}{\partial r}\right) \right],
\label{psidef}
\eeq
where $K(t,r)$ and $H_2(t,r)$ are the perturbations, and the Zerilli \cite{ze70a} normalisation is used for $\Psi_l $. The wave equation is given by the operator $\mathcal{Z}$ acting on the wave function  

\beq
\fl\qquad\qquad\mathcal{Z}\Psi^l(t,r) = {\partial^2_{r^*}} \Psi^l(t,r) - 
{\partial^2_{t}}\Psi^l(t,r) -V^{l}(r)\Psi^l (t,r) = S^{l}(t,r)~,
\label{eq:rwz*}
\eeq
where $r^* = r + 2M\ln (r/2M - 1)$ is the tortoise coordinate and the potential $V^{l}(r)$ is

\beq
V^{l}(r) = \left (1 - \frac{2M}{r}\right )
\frac {2\lambda^{2}(\!\lambda\!+\!1\!) r^{3}\!\!+\! 6 \lambda^{2}Mr^{2} \!+\! 18\lambda M^{2}r\!+\! 18M^{3}} {r^{3}(\lambda r \!+\! 3M)^2}~,
\eeq
being $ \lambda = 1/2(l - 1)(l + 2) $. The source $S^{l}(t,r)$ includes the derivative of the Dirac distribution (the latter appear in the process of forming the wave equation out of the ten linearised Einstein equations)

\[
S^{l} = \frac{2( r- 2M ) 
\kappa }{r^2(\lambda +1)(\lambda r+3M)}
\times 
\]
\beq
\fl
\left \{
\frac{r (r-2M )}{2U^0}
\delta^{\prime }\left[r-r_u(t)\right]- 
\left [ \frac{r(\lambda + 1)- 3M}{2U^0} - \frac{3MU^0(r-2M)^2}{r(\lambda r+3M )} \right ]
\delta \left[r-r_u(t)\right]
\right \}~,
\eeq
{ $U^0 = E/(1 - 2M/r_u)$ being the time component of the 4-velocity, $E=\sqrt{1 - 2M/r_{u0}}$ the conserved energy per unit mass,} and $\kappa = 4 m\sqrt{(2l+1)\pi}$. The geodesic in the unperturbed SD metric $z_u(\tau)=\{t_u(\tau),r_u(\tau),\theta_u(\tau),\phi_u(\tau)\}$ assumes different forms according to the 
initial conditions.   
For radial infall of a particle starting
from rest at finite distance $r_{u0}$, $r_{u}(t)$ is the inverse function in coordinate time $t$ of the trajectory in the background field, corresponding to the geodesic in proper time $\tau$ (u stands for unperturbed)

\[
\fl\frac{t(r_u)}{2M}= \sqrt{1-\frac{2M}{r_{u0}}}\sqrt{1-\frac{r_u}{r_{u0}}}\left(\frac{r_{u0}}{2M}\right)\left(\frac{r_u}{2M}\right)^{1/2} +2{\rm{arctanh}}\left(\frac{\sqrt{\ds\frac{2M}{r_u}-\frac{2M}{r_{u0}}}}{\sqrt{\ds 1-\frac{2M}{r_{u0}}}}\right) +
\]
\beq
\sqrt{1-\frac{2M}{r_{u0}}}\left(1+\frac{4M}{r_{u0}}\right)\left(\frac{r_{u0}}{2M}\right)^{3/2}
\arctan\left(\sqrt{\frac{r_{u0}}{r}-1}\right)~.
\eeq

The expressions above correspond to those in \cite{sp11}, where some of the errors of previously published literature on radial fall are indicated. 

\section{Jump conditions}

From the visual inspection of the Zerilli wave equation (\ref{eq:rwz*}), it is evinced that the wave function $\Psi$ is of $C^{-1}$ continuity class (the second derivative of the wave function is proportional to the first derivative of the Dirac distribution, in itself a $C^{-3}$ class element). Thus, the wave function and its derivatives can be written as (the $l$ index is  dropped henceforth for simplicity of notation)

\bea
\fl\Psi &=& \Psi^+ \Theta_1+\Psi^-\Theta_2~,\label{psi}\\
\fl\Psi_{,r} &=&  \Psi^+_{,r}\Theta_1 + \Psi^-_{,r} \Theta_2 + \left(\Psi^+ -\Psi^-\right) \delta~,\label{psir}\\
\fl\Psi_{,t} &=&  \Psi^+_{,t}\Theta_1  +  \Psi^-_{,t} \Theta_2- \dot{r}_u\left(\Psi^+ -\Psi^-\right)\delta~,\label{psit}\\
\fl\Psi_{,rr} &=&  \Psi^+_{,rr}\Theta_1  +  \Psi^-_{,rr} \Theta_2 + 2\left( \Psi^+_{,r} - \Psi^-_{,r} \right) \delta + \left(\Psi^+ - \Psi^-\right) \delta'~, \label{psirr}\\
\fl\Psi_{,tt} &=& \Psi^{+}_{,tt}\Theta_1  +  \Psi^{-}_{,tt} \Theta_2-2\dot{r}_u\left(\Psi^+_{,t} -\Psi^-_{,t}\right)\delta - \ddot{r}_u\left(\Psi^+ -\Psi^-\right)\delta +\dot{r}_u^2\left(\Psi^+ -\Psi^-\right)\delta'
,\label{psitt}\\
\fl\Psi_{,tr} &=& \Psi^+_{,tr}\Theta_1  +  \Psi^-_{,tr} \Theta_2  +  \left(\Psi^+_{,t} -\Psi^-_{,t}\right) \delta-\dot{r}_u\left(\Psi^+_{,r} -\Psi^-_{,r}\right) \delta- \dot{r}_u\left(\Psi^+ -\Psi^-\right)\delta'~~\label{psitr},
\eea
where in shortened notation $\Theta_1 = \Theta\left[r-r_u(t)\right]$, and 
$\Theta_2 = \Theta\left[r_u(t) - r \right]$ are two Heaviside step distributions, while 
$\delta = \delta\left[r-r_u(t)\right ]$ and $\delta' = \delta'\left[r-r_u(t)\right]$ { are the Dirac delta - and its derivative - distributions}. The dot and the prime indicate time and space derivatives, respectively. 

\subsection{Jump conditions from the wave equation}

For the computation of { back-action} effects, we need first order derivatives of the perturbations and thus third order wave function derivatives. To this end, we operate directly on the wave equation, Eq. \ref{eq:rwz*}. The source term is 
cast in the following form

\beq
S(t,r) =  G(t,r)\delta + F(t,r)\delta' = \tilde{G}_{r_u(t)}\delta + F_{r_u(t)}\delta'~, 
\label{eq:rwz*source}
\eeq
where $\tilde{G}_{r_u(t)} =  G_{r_u(t)}- F'_{r_u(t)}$ and one of the properties of the Dirac delta distribution, namely 
$\phi(r)\delta'\left[r-r_u(t)\right]=\phi_{r_u(t)} \delta'\left[r-r_u(t)\right] - \phi'_{r_u(t)} 
\delta\left[r-r_u(t)\right]${ ,} has been used at the position of the particle. The determination of the jump conditions imposes { the transformation of} Eq. \ref{eq:rwz*} into the corresponding equation in 
(r,t) domain (the tortoise coordinate can't be inverted). Turning to the $r$ variable, we get $\left( f = 1 - 2M/r\right)$ 

\bea
\fl\qquad {\partial^2_{r^*}}\Psi & = & ff'\partial_r\Psi+f^2{\partial^2_{r}}\Psi \nonumber \\
\fl\qquad & = & \left[ff'\Psi^+_{,r}+f^2\Psi_{,rr}^+\right]\Theta_1+\left[ff'\Psi^-_{,r}+f^2\Psi_{,rr}^-\right]\Theta_2 +ff'\left(\Psi^+-\Psi^-\right)\delta\nonumber\\
\fl\qquad &\ &+2f^2\left(\Psi_{,r}^+-\Psi_{,r}^-\right)\delta+f^2\left(\Psi^+-\Psi^-\right)\delta'~,\\
\fl\qquad\nonumber\\
\fl\qquad \partial^2_{t}\Psi & = & \Psi_{,tt}^+\Theta_1 +\Psi_{,tt}^+\Theta_2- 2\dot{r}_u\partial_t \left(\Psi^+-\Psi^-\right)\delta -\ddot{r}_u\left(\Psi^+-\Psi^-\right)\delta \nonumber\\
\fl\qquad &\ &+ \dot{r}_u^2\left(\Psi^+-\Psi^-\right)\delta'~,\\
\fl\qquad\nonumber\\
\fl\qquad V \Psi & = & V \Psi^+\Theta_1 + V \Psi^-\Theta_2~.
\eea

The notation $[\Psi]$ stands for the difference $\left (\Psi^+ - \Psi^- \right )_{r_u}$ and a likewise notation is used for the derivatives at the point $r_u$. Equating the coefficients of $\delta'$, and owing to the above mentioned property of the delta derivative for which $\left(\Psi^+-\Psi^-\right)\delta' = 
\left[\Psi\right]\delta'-\left[\Psi_{,r}\right]\delta$, we get the jump condition for $\Psi$

\beq
\left[\Psi\right]=\frac{1}{f_{r_u}^2-\dot{r}_u^2}F_{r_u}~.
\eeq

\noindent Equating the coefficients of $\delta$, we get the jump condition on the space derivative

\beq
\left[\Psi_{,r}\right]=\frac{1}{f_{r_u}^2-\dot{r}_u^2}\left[\tilde{G}_{r_u}+\left(f_{r_u}f_{r_u}'-\ddot{r}_u\right)\left[\Psi\right]-2\dot{r}_u\frac{d}{dr_u}\left[\Psi\right]\right]~,
\eeq

\noindent and therefore the jump condition on the first time derivative

\beq
\left[\Psi_{,t}\right]=\dot{r}_u\frac{d}{dr_u}\left[\Psi\right]-\dot{r}_u\left[\Psi_{,r}\right]~.
\eeq

\noindent Since $\mathcal{Z}\Psi^\pm=0$, the coefficients of $\Theta_1$ and $\Theta_2$ ought to be equal. We thus obtain

\beq
\left[\Psi_{,tt}\right] - f_{r_u}f_{r_u}'\left[\Psi_{,r}\right] - f_{r_u}^2\left[\Psi_{,rr}\right] +V_{r_u}\left[\Psi\right]=0~,
\eeq

\noindent which is an equation with two unknowns. We circumvent the difficulty by using i) the commutativity of the derivatives, $\left[\Psi_{,tr}\right]=\left[\Psi_{,rt}\right]$, ii) the transformation $d/dt = \dot{r}_u d/dr_u$, and write

\bea
\fl\qquad\left[\Psi_{,tt}\right]&=\frac{d}{dt}\left[\Psi_{,t}\right]-\dot{r}_u\left[\Psi_{,tr}\right]=\frac{d}{dt}\left[\Psi_{,t}\right]-\dot{r}_u\left\{\frac{d}{dt}\left[\Psi_{,r}\right]-\dot{r}_u\left[\Psi_{,rr}\right]\right\} = \nonumber\\
\fl\qquad&=\dot{r}_u\frac{d}{dr_u}\left[\Psi_{,t}\right]-\dot{r}_u^2\frac{d}{dr_u}\left[\Psi_{,r}\right]+\dot{r}_u^2\left[\Psi_{,rr}\right]~.
\eea

\noindent The jump condition on the second space derivative can now be expressed by

\beq
\fl\qquad
\left[\Psi_{,rr}\right]=\frac{1}{f_{r_u}^2-\dot{r}_u^2}\left\{\dot{r}_u\frac{d}{dr_u}\left[\Psi_{,t}\right]-\dot{r}_u^2\frac{d}{dr_u}\left[\Psi_{,r}\right]-f_{r_u}f_{r_u}'\left[\Psi_{,r}\right]+V_{r_u}\left[\Psi\right]\right\}~.
\eeq

\noindent The other second derivatives are obtained by

\bea
&\left[\Psi_{,tr}\right]=\left[\Psi_{,rt}\right]=\frac{d}{dt}\left[\Psi_{,r}\right]-\dot{r}_u\left[\Psi_{,rr}\right]~,\\
&\left[\Psi_{,tt}\right]=\frac{d}{dt}\left[\Psi_{,t}\right]-\dot{r}_u\left[\Psi_{,tr}\right]~.
\eea

\noindent For the third order derivatives, we derive the wave equation with respect to $r$ and obtain

\[
\left[\Psi_{,rrr}\right]=\frac{1}{\dot{r}_u^2-f_{r_u}^2}
\Bigg\{\dot{r}_u^2\frac{d}{dr_u}\left[\Psi_{,rr}\right] -\dot{r}_u\frac{d}{dr_u}\left[\Psi_{,rt}\right]
\]
\beq
+\left(f_{r_u}'^2+f_{r_u}f_{r_u}''-V_{r_u}\right)\left[\Psi_{,r}\right]+3f_{r_u}f_{r_u}'\left[\Psi_{,rr}\right]-V_{r_u}'\left[\Psi\right]\Bigg \}~,
\eeq

\noindent while deriving with respect to $t$, we obtain

\beq
\fl
\left[\Psi_{,ttt}\right]=\frac{\dot{r}_u^2}{\dot{r}_u^2-f_{r_u}^2}\left\{f_{r_u}^2\frac{d}{dr_u}\left[\Psi_{,rt}\right]-\dot{r}_u^{-1}f_{r_u}^2\frac{d}{dr_u}\left[\Psi_{,tt}\right]+f_{r_u}f_{r_u}'\left[\Psi_{,rt}\right]-V_{r_u}\left[\Psi_{,t}\right]\right\}~,
\eeq

\bea
&\left[\Psi_{,ttr}\right]=\left[\Psi_{,trt}\right]=\left[\Psi_{,rtt}\right]=\frac{d}{dr_u}\left[\Psi_{,tt}\right]-\dot{r}_u^{-1}\left[\Psi_{,ttt}\right]~,\\
&\left[\Psi_{,trr}\right]=\left[\Psi_{,rtr}\right]=\left[\Psi_{,rrt}\right]=\frac{d}{dr_u}\left[\Psi_{,tr}\right]-\dot{r}_u^{-1}\left[\Psi_{,ttr}\right]~,\\
&\left[\Psi_{,rrr}\right]=\frac{d}{dr_u}\left[\Psi_{,rr}\right]-\dot{r}_u^{-1}\left[\Psi_{,trr}\right]~.
\eea

\noindent Finally, we similarly proceed for the fourth derivatives

\[
\!\!\!\!\!\!\!\!\!\!\!\!\left[\Psi_{,tttt}\right]=\frac{\dot{r}_u^2}{\dot{r}_u^2-f_{r_u}^2} \times
\]
\beq
\!\!\!\!\!\!\!\!\!\!\!\!\left\{f_{r_u}^2\frac{d}{dr_u}\left[\Psi_{,ttr}\right]-\dot{r}_u^{-1}f_{r_u}^2\frac{d}{dr_u}\left[\Psi_{,ttt}\right]+f_{r_u}f_{r_u}'\left[\Psi_{,ttr}\right]-V_{r_u}\left[\Psi_{,tt}\right]\right\},
\eeq

\bea
&\!\!\!\!\!\!\!\!\!\!\!\!\!\!\left[\Psi_{,tttr}\right]=\left[\Psi_{,ttrt}\right]=\left[\Psi_{,trtt}\right]=\left[\Psi_{,rttt}\right]\frac{d}{dr_u}\left[\Psi_{,ttt}\right]-\dot{r}_u^{-1}\left[\Psi_{,tttt}\right]~,\\
&\!\!\!\!\!\!\!\!\!\!\!\!\!\!
\left[\Psi_{,ttrr}\right]=\left[\Psi_{,trtr}\right]= \left[\Psi_{,trrt}\right]= \left[\Psi_{,rttr}\right]\left[\Psi_{,rtrt}\right]= \left[\Psi_{,rrtt}\right] = \nonumber \\
& ~~~= \frac{d}{dr_u}\left[\Psi_{,ttr}\right]-\dot{r}_u^{-1}\left[\Psi_{,tttr}\right]~,\\
&\!\!\!\!\!\!\!\!\!\!\!\!\!\!\!\left[\Psi_{,trrr}\right]= \left[\Psi_{,rtrr}\right]=\left[\Psi_{,rrtr}\right]\left[\Psi_{,rrrt}\right]\frac{d}{dr_u}\left[\Psi_{,trr}\right]-\dot{r}_u^{-1}\left[\Psi_{,ttrr}\right]~,\\
&\!\!\!\!\!\!\!\!\!\!\!\!\!\!\left[\Psi_{,rrrr}\right]=\frac{d}{dr_u}\left[\Psi_{,rrr}\right]-\dot{r}_u^{-1}\left[\Psi_{,rrrt}\right]~.
\eea

\subsubsection{Jump conditions in explicit form.}

We list hereafter the jump conditions in explicit form.

\paragraph{Jump conditions}

\beq
[\Psi]  = 
\frac{\kappa E r_u}{(\lambda +1) (3 M+\lambda   r_u)}
\label{jump-psi} 
\eeq

\paragraph{First derivative jump conditions}
\beq
[\Psi_{,t}]  =  
-\frac{\kappa E r_u \dot{r}_{u}}{(2 M - r_u) (3
   M+\lambda r_u)}
\label{jump-psit} 
\eeq

\beq
[\Psi_{,r}] = 
\frac{\kappa E \left[6 M^2+3 M \lambda  r_u+\lambda 
   (\lambda +1) r_u^2\right]}{(\lambda +1) (2 M-r_u) (3 M+\lambda  r_u)^2}
\label{jump-psir}
\eeq

\paragraph{Second derivative jump conditions}
\beq
\fl [\Psi_{,rr}] = 
-\frac{\kappa E \left[3 M^3 (5 \lambda -3)+6 M^2 \lambda (\lambda -3)
    r_u+3 M \lambda ^2
(\lambda -1)    r_u^2-2 \lambda ^2 (\lambda +1)
   r_u^3\right]}{(\lambda +1) (2M-r_u)^2 (3 M+\lambda  r_u)^3}
\label{jump-psirr}
\eeq

\beq
[\Psi_{,tr}] = 
\frac{\kappa E \left(3 M^2+3 M \lambda 
   r_u - \lambda  r_u^2\right)\dot{r}_{u}}{(2M - r_u)^2 (3 M+\lambda  r_u)^2}
\label{jump-psitr} 
\eeq

\beq
[\Psi_{,tt}] = -\frac{\kappa\,E\,M}{{r_u}^{2}\,\left( 3\,M+r_u\,\lambda\right) }
\label{jump-psitt} 
\eeq

\paragraph{Third derivative jump conditions}

\bea
\fl\left[\Psi_{,rrr}\right]=&
\frac{\kappa E}{r_u\left(  \lambda+1\right) {\left( 2M-r_u\right) }^{3}{\left( 3M+r_u \lambda\right) }^{4}}\bigg[81\left( \lambda+1\right) {M}^{5}+9r_u\left( 19\lambda^{2}+18{E}^{2}\lambda + \right.\nonumber\\
\fl&\left. 3\lambda+18{E}^{2}\right) {M}^{4}+9r_u^{2}\lambda\left(7\lambda^{2}+24{E}^{2}\lambda-14\lambda
+24{E}^{2}+3\right) {M}^{3}+3r_u^{3}\lambda^{2}\left( 7 \lambda ^2\right.\nonumber\\
\fl&\left.+36{E}^{2}\lambda-11\lambda+36{E}^{2}+18\right){M}^{2}+3r_u^{4}\lambda^{3}\left( 8{E}^{2}\lambda-7\lambda+8{E}^{2}-1\right) M + \nonumber\\
\fl& 2r_u^{5}\lambda^{3}\left( \lambda+1\right) \left( {E}^{2}\lambda+3\right) \bigg]
\label{jump-psirrr}
\eea

\bea
\fl\left[\Psi_{,trr}\right]= &  
\frac{-\kappa E\dot{r}_u}{r_u{\left( 2M-r_u\right) }^{3}{\left( 3M+r_u \lambda\right) }^{3}}\bigg[27{M}^{4}+6r_u\left( 5\lambda+9{E}^{2}-3\right) {M}^{3}+3r_u^{2}\lambda \left( 5\lambda + \right.\nonumber\\
\fl&\left. 18{E}^{2}-6\right) {M}^{2}+6r_u^{3}\lambda^{2}\left( 3{E}^{2}-2\right) M+2r_u^{4}\lambda^{2}\left( {E}^{2}\lambda+1\right) \bigg]
\eea

\bea
\fl\left[\Psi_{,ttr}\right]=&
\frac{\kappa E}{r_u^{3}\left( 2M-r_u\right) {\left( 3M+r_u \lambda\right) }^2}\bigg[39{M}^{3}+9r_u\left( 3\lambda+2{E}^{2}-2\right) {M}^{2}+r_u^{2}\lambda\left( 4\lambda + \right. \nonumber\\
\fl& \left. 12{E}^{2}-13\right) M+2r_u^{3}\lambda^{2}\left( E^2-1\right) \bigg]
\eea

\beq
\fl\left[\Psi_{,ttt}\right]=\frac{-\kappa E\dot{r}_u}{r_u^{3}\left(2M-r_u\right)\left( 3M+r_u \lambda\right)}\bigg[9{M}^{2}+2r_u\left( 2\lambda+3{E}^{2}-2\right) M+2r_u^{2}\lambda \left( E^2-1\right)\bigg]
\eeq

\paragraph{ Fourth derivative jump conditions} 
\bea
\fl\left[\Psi_{,rrrr}\right]=& \frac{-3\kappa E}{r_u^2\left(  \lambda+1\right) {\left( 2M-r_u\right) }^{4}{\left( 3M+r_u \lambda\right) }^{5}}\bigg[567\left( \lambda+1\right) {M}^{7}+162r_u\left( \lambda+1\right) \left( 6\lambda\right.\nonumber\\
\fl&\left.+16{E}^{2}-5\right) {M}^{6}+6r_u^{2}\left( 139\lambda^{3}+738{E}^{2}\lambda^{2}-123\lambda^{2}+162{E}^{4}\lambda+441{E}^{2}\lambda\right.\nonumber\\
\fl&\left.-171\lambda+162{E}^{4}-297{E}^{2}+27\right) {M}^{5}+12r_u^{3}\lambda\left( 21\lambda^{3}+252{E}^{2}\lambda^{2}-85\lambda^{2}+\right.\nonumber\\
\fl&\left.135{E}^{4}\lambda-24\lambda+135{E}^{4}-252{E}^{2}+18\right) {M}^{4}+3r_u^{4}\lambda^{2}\left( 21\lambda^{3}+344{E}^{2}\lambda^{2}-\right.\nonumber\\
\fl&\left.95\lambda^{2}+360{E}^{4}\lambda-340{E}^{2}\lambda+100\lambda+360{E}^{4}-684{E}^{2}+24\right) {M}^{3}+2r_u^{5}\lambda^{3}\!\cdot\nonumber\\
\fl&\left( 88{E}^{2}\lambda^{2}-47\lambda^{2}+180{E}^{4}\lambda-260{E}^{2}\lambda+25\lambda+180{E}^{4}-348{E}^{2}-24\right) {M}^{2}\nonumber\\
\fl&+2r_u^{6}\lambda^{4}\left( 6{E}^{2}\lambda^{2}+30{E}^{4}\lambda-53{E}^{2}\lambda+23\lambda+30{E}^{4}-59{E}^{2}+11\right) M+\nonumber\\
\fl&4r_u^{7}\lambda^{4}\left( \lambda+1\right) \left( {E}^{4}\lambda-2{E}^{2}\lambda-2\right)\bigg]
\eea

\bea
\fl\left[\Psi_{,trrr}\right]=&\frac{3\kappa E\dot{r}_u}{r_u^2{\left( 2M-r_u\right) }^{4}{\left( 3M+r_u \lambda\right) }^{4}}\bigg[135{M}^{6}+27r_u\left( 7\lambda+32{E}^{2}-6\right) {M}^{5}+3r_u^{2}\!\cdot\nonumber\\
\fl&\left( 35\lambda^{2}+396{E}^{2}\lambda-75\lambda+108{E}^{4}-144{E}^{2}+18\right) {M}^{4}+r_u^{3}\lambda\left( 35\lambda^{2}+\right.\nonumber\\
\fl&\left.612{E}^{2}\lambda-120\lambda+432{E}^{4}-594{E}^{2}+72\right) {M}^{3}+r_u^{4}\lambda^{2}\left( 140{E}^{2}\lambda-45\lambda+\right.\nonumber\\
\fl&\left.216{E}^{4}-306{E}^{2}+36\right) {M}^{2}+2r_u^{5}\lambda^{3}\left( 6{E}^{2}\lambda+24{E}^{4}-35{E}^{2}+9\right) M+\nonumber\\
\fl&2r_u^{6}\lambda^{3}\left( 2{E}^{4}\lambda-3{E}^{2}\lambda-1\right)\bigg]
\eea

\bea
\fl\left[\Psi_{,ttrr}\right]=&\frac{-\kappa E}{r_u^{4}{\left( 2M-r_u\right) }^2{\left( 3M+r_u \lambda\right) }^{3}}\bigg[1431{M}^{5}+6r_u\left( 251\lambda+234{E}^{2}-210\right) {M}^{4}+\nonumber\\
\fl&9r_u^{2}\left( 59\lambda^{2}+160{E}^{2}\lambda-148\lambda+36{E}^{4}-66{E}^{2}+30\right) {M}^{3}+6r_u^{3}\lambda\left( 10\lambda^{2}+\right.\nonumber\\
\fl&\left.82{E}^{2}\lambda-79\lambda+54{E}^{4}-102{E}^{2}+48\right) {M}^{2}+2r_u^{4}\lambda^{2}\left( 28{E}^{2}\lambda-27\lambda+54{E}^{4}\right.\nonumber\\
\fl&\left.-105{E}^{2}+52\right) M+12r_u^{5}\lambda^{3}{\left( E^2-1\right) }^{2}\bigg]
\eea

\bea
\fl\left[\Psi_{,tttr}\right]=&\frac{\kappa E\dot{r}_u}{r_u^{4}\left( 2M-r_u\right)^2{\left( 3M+r_u \lambda\right) }^2}\bigg[243{M}^{4}+3r_u\left( 61\lambda+132{E}^{2}-64\right) {M}^{3}+3r_u^{2}\!\cdot\nonumber\\
\fl&\left( 12\lambda^{2}+92{E}^{2}\lambda-49\lambda+36{E}^{4}-48{E}^{2}+12\right) {M}^{2}+2r_u^{3}\lambda\left( 24{E}^{2}\lambda-15\lambda+\right.\nonumber\\
\fl&\left.36{E}^{4}-51{E}^{2}+14\right) M+6r_u^{4}\lambda^{2}\left( E^2-1\right) \left( 2{E}^{2}-1\right) \bigg]
\eea

\bea
\fl\left[\Psi_{,tttt}\right]=&\frac{-\kappa E}{{r_u^{6}\left( 3M+r_u \lambda\right) }}\bigg[189{M}^{3}+2r_u\left( 36\lambda+84{E}^{2}-77\right) {M}^{2}+6r_u^{2}\left( E^2-1\right)\left( 10\lambda+\right.\nonumber\\
\fl&\left.6{E}^{2}-5\right) M+12r_u^{3}\lambda{\left( E^2-1\right) }^{2}\bigg]
\eea

{ 
While heuristic arguments \cite{lo00, lona09} have been put forward to show that, for radial fall in the RW gauge, even metric perturbations belong to the $C^0$ continuity class at the position of th particle, in} \cite{aosp11a, aosp11b} we have provided an analysis {\it vis \`{a} vis} the jump conditions that the wave function and its { (first and second) } derivatives have to satisfy for guaranteeing the continuity of the perturbations at the position of the particle. Therein, we have derived the same jump conditions (\ref{jump-psi} - \ref{jump-psitr}) from the inverse relations (expressions giving the perturbations as function of the wave function and its derivatives) by fulfilment of the continuity conditions (equal coefficients for the two Heaviside distributions, and null coefficients for the Dirac distribution and its derivative). 

\section{The algorithm} 

The integration method considers cells belonging to two groups 
for cells never crossed by the world line, the integrating method may be drawn by previous approaches explored by Lousto \cite{lo05} and Haas \cite{ha07}, whereas for cells crossed by a particle, we propose a new algorithm. The grid is in the $r^*,t$ domain. 

Initial conditions require knowledge of the situation prior to $t=0$. At fourth order, the wave function may be Taylor-expanded around $t=0$. For the boundary conditions, simplicity suggests a sufficiently huge grid to avoid unwanted reflections. 

\subsection{Empty cells}

\begin{figure}[H]
\begin{center}
\includegraphics[width=10cm]{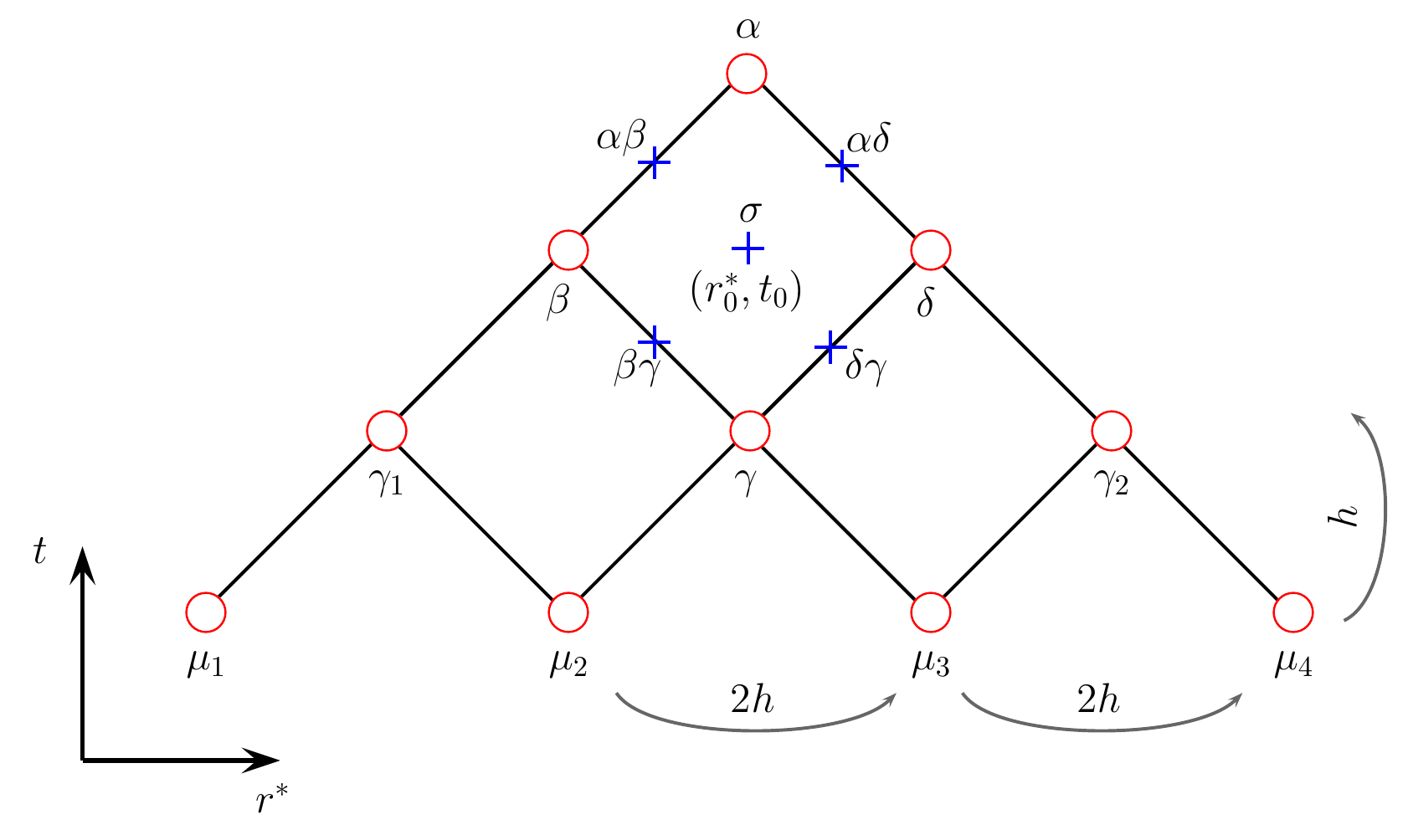}
\caption{\small Set of points (circles and crosses) used for the integration of the $V\Psi { = g}$ term in the vacuum case. 
The crosses don't overlap with grid nodes; thus the field $g$ at these points, Eqs. (\ref{eq.sumg},\ref{eq.g0}), is approximated by the field at the nodes on the past light cone of the grid node $\alpha$.}
\label{fig.intgdA}
\end{center}
\end{figure}

{ Empty cells are those cells which are not crossed by the particle. In this case}, the cell upper point is obtained by performing an integration of the wave equation over the entire surface $ A$ of the cell, identified by the nodes $\alpha,\beta,\gamma,\delta$. We briefly recall the algorithm used by Haas \cite{ha07}. Therein, the sole numerical computation to be carried out is represented by the product of the potential term and the wave function $V\Psi { = g}$. It is performed via a double Simpson integral, using points of the past light cone of the upper node $\alpha$, Fig. \ref{fig.intgdA}. 
We set $g_q=g(r^*_q,t_q)=V(r_q)\Psi(r^*_q,t_q)$, $V_q=V(r_q)$ and $\Psi_q=\Psi(r^*_q,t_q)$, where $q$ is one of the points shown in Fig. \ref{fig.intgdA}. The increment $h$ is defined as $h=\frac{1}2\Delta r^*=\frac{1}2\Delta t$ where $\Delta r^*$ is the spatial step and $\Delta t$ is the time step. 

We have

\beq
\fl\int\!\!\!\!\int_{Cell} gdA=\left(\frac{h}{3}\right)^2\Big[g_\alpha+g_\beta+g_\gamma+g_\delta+4\left(g_{\beta\gamma}+g_{\alpha\beta}+g_{\delta\gamma}+g_{\alpha\delta}\right)+16g_\sigma\Big]+{\cal O}(h^6)~,
\label{gda}
\eeq
where the sum of the intermediate terms between nodes is given by
\bea
\fl g_{\beta\gamma}+g_{\alpha\beta}+g_{\delta\gamma}+g_{\alpha\delta} &=& 2V_\sigma\Psi_\sigma\Big[1-\frac{1}2\left(\frac{h}2\right)^2V_\sigma\Big] + V_{\beta\gamma}\Psi_\beta\Big[1-\frac{1}2\left(\frac{h}2\right)^2V_{\beta\gamma}\Big] + \nonumber\\
\fl &\ & V_{\delta\gamma}\Psi_\delta\Big[1-\frac{1}2\left(\frac{h}2\right)^2V_{\delta\gamma}\Big] + \frac{1}2\Big[V_{\beta\gamma}-2V_\sigma+V_{\delta\gamma}\Big]\left(\Psi_\beta+\Psi_\delta\right)\nonumber\\
\fl &\ &+{\cal O}(h^4)~.
\label{eq.sumg}
\eea
The last intermediate term $g_\sigma$ in Eq. \ref{gda} is evaluated using given nodes in the past light cone of $\alpha$, Fig. \ref{fig.intgdA}
\beq
\fl\qquad g_\sigma=\frac{1}{16}\Big[8g_\beta+8g_\gamma+8g_\delta-4g_{\gamma_1}-4g_{\gamma_2}+g_{\mu_1}-g_{\mu_2}
- g_{\mu_3}+g_{\mu_4}\Big]+{\cal O}(h^4)~.
\label{eq.g0}
\eeq

\noindent For the differential operators, an exact integration simply leads to

\beq
\fl\qquad\qquad\int\!\!\!\!\int_{Cell} \left({\partial^2_{r^*}}-\partial^2_t\right)\Psi(r^*,t)dA=-4\Big[\Psi_\alpha-\Psi_\beta+\Psi_\gamma-\Psi_\delta\Big]~.
\eeq

\noindent Finally, we get
\bea
\fl\qquad \Psi_\alpha=-\Psi_\gamma &+& \Psi_\beta\Big[1 - \frac{1}4\left(\frac{h}2\right)^2\left(V_\sigma+V_\beta\right) + \frac{1}{16}\left(\frac{h}2\right)^4V_\sigma\left(V_\sigma+V_\beta\right)\Big] \nonumber\\
\fl\qquad &+& \Psi_\delta\Big[1 - \frac{1}4\left(\frac{h}2\right)^2\left(V_\sigma+V_\delta\right) + \frac{1}{16}\left(\frac{h}2\right)^4V_\sigma\left(V_\sigma+V_\delta\right)\Big] \nonumber\\
\fl\qquad &-& \left(\frac{h}2\right)^2\Big[1-\frac{1}4\left(\frac{h}2\right)^2V_\sigma\Big]\Big[g_{\beta\gamma}+g_{\alpha\beta}+g_{\delta\gamma}+g_{\alpha\delta}+4g_\sigma\Big]~.
\eea

For cells adjacent to cells crossed by the particle, the requirement of good accuracy suggests a different dealing for the
computation of $g_\sigma$, { since the past light cone of an adjacent cell} can cross the path of the particle. In such a case, $g_\sigma$ is approximated by non-centred spatial finite difference expressions \cite {ha07}. 

\subsection{Cells crossed by the world line}
\begin{figure}[H]
\begin{center}
\includegraphics[width=14cm]{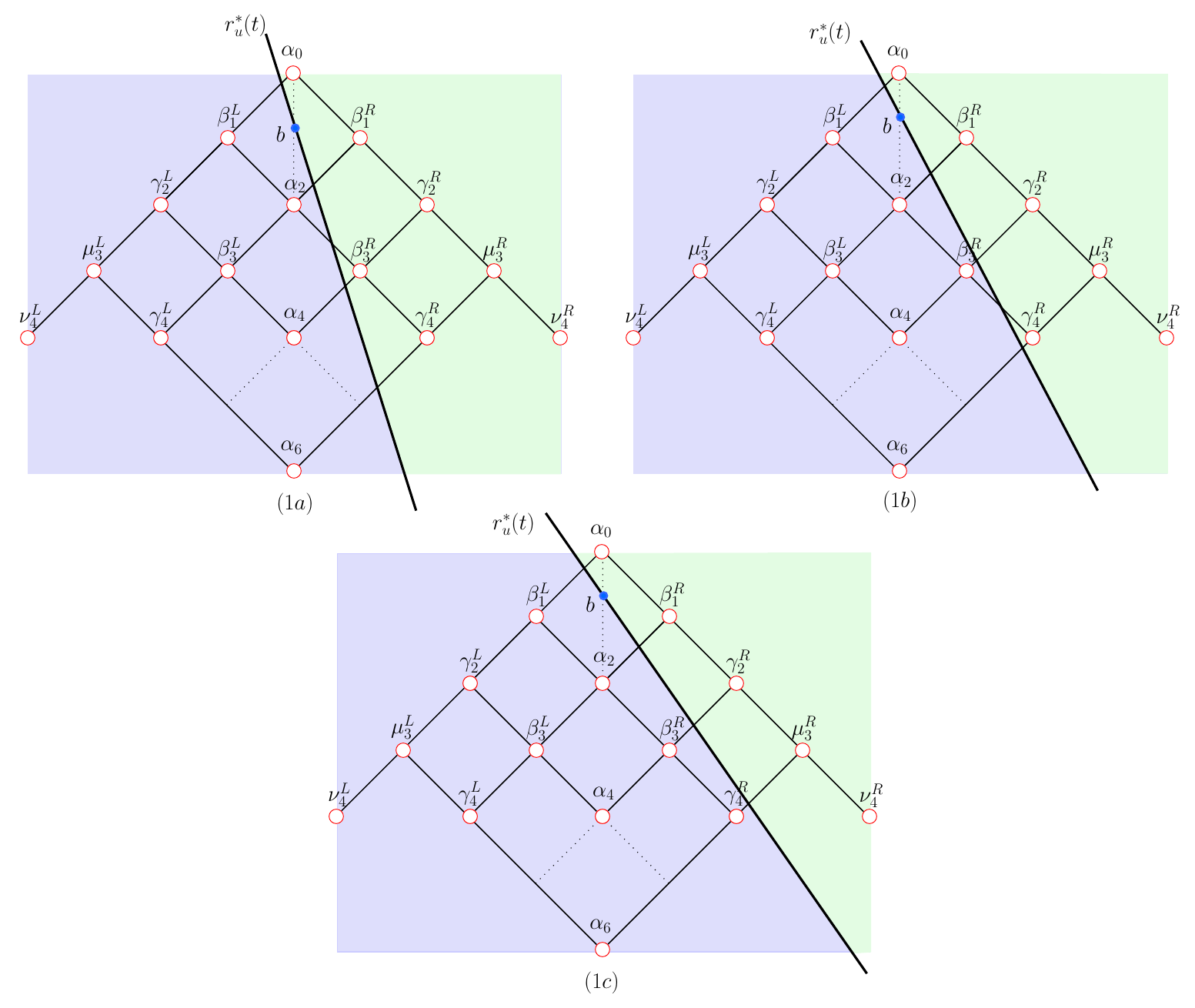}
\end{center}
\caption{\small{The three sub-cases for which the particle enters through the $[\alpha_2\beta^R_1]$ side and leaves through the $[\alpha_0\beta^L_1]$ side. The elimination of the $\Psi^-_b$ derivatives demands, Eq. \ref{scheme1_with_der}, the utilisation of fifteen points, represented by circles, in the light cone of $\alpha_0$. Numerical efficiency suggests that the points are taken at both left and right sides of the $r_u^*(t)$ trajectory. In the three cases, the particle crosses the line $\left[\alpha_0 \alpha_2 \right]$ at the point $b$. The background distinguishes two zones: one where  $\Psi(r^*\!<\!r_u^*(t),t)=\Psi^-(r^*,t)$, the other where $\Psi(r^*\!>\!r_u^*(t),t)=\Psi^+(r^*,t)$, the path $r_u^*(t)$ representing the separation between the two zones.}}
\label{fig.case1}
\end{figure}

\begin{figure}[H]
\begin{center}
\includegraphics[width=14cm]{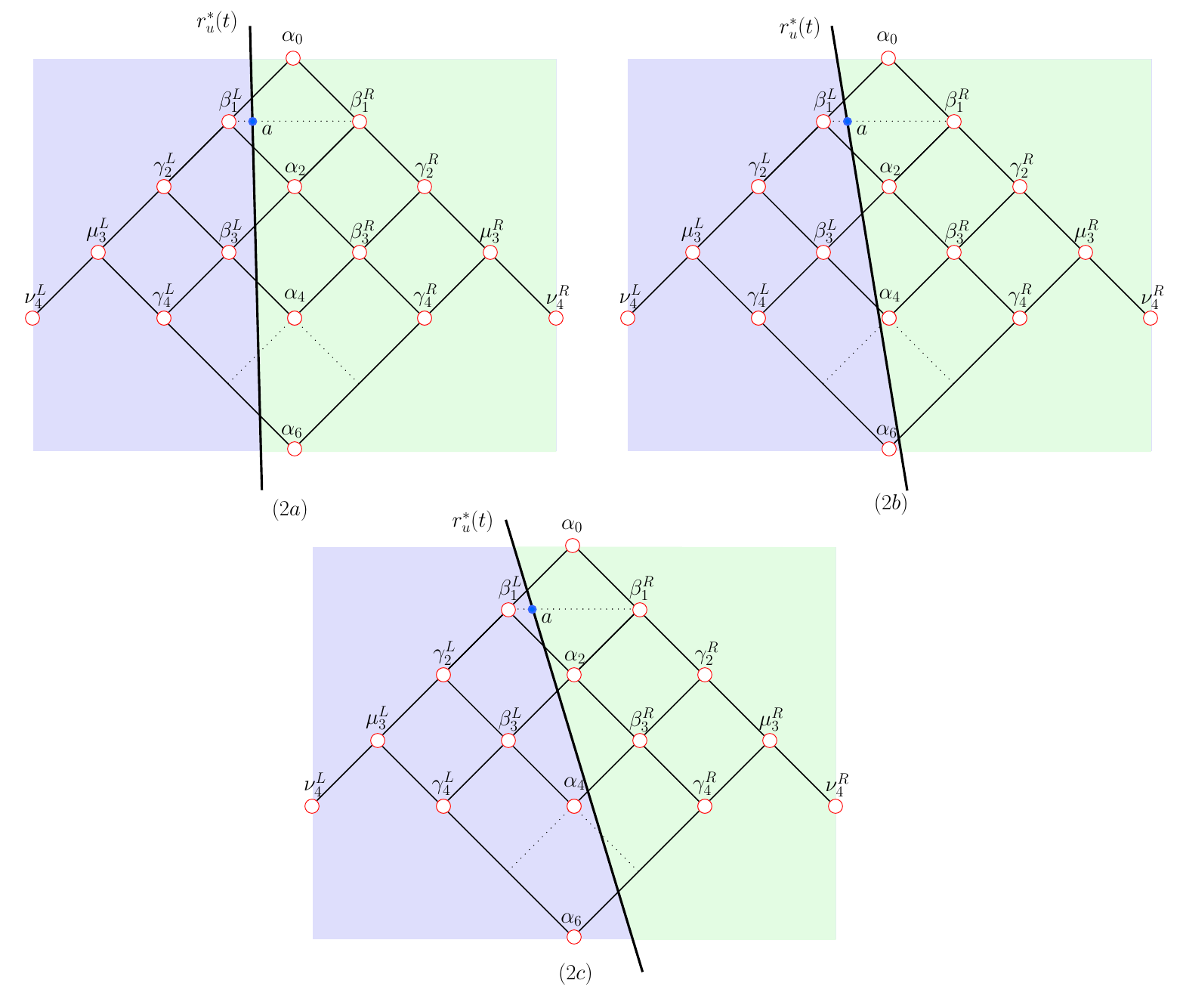}
\end{center}
\caption{\small{The three sub-cases for which the particle enters through the $[\alpha_2\beta^L_1]$ side and leaves through the $[\alpha_0\beta^L_1]$ side. The elimination of the $\Psi^-_a$ derivatives demands the utilisation of fifteen points, represented by circles, in the light cone of $\alpha_0$. Numerical efficiency suggests that the points are taken at both left and right sides of the $r_u^*(t)$ trajectory. In the three cases, the particle crosses the line $\left[\beta^L_1 \beta^R_1 \right]$ at the point 
$a$. 
The background distinguishes two zones: one where  $\Psi(r^*\!<\!r_u^*(t),t)=\Psi^-(r^*,t)$, the other where 
$\Psi(r^*\!>\!r_u^*(t),t)=\Psi^+(r^*,t)$, the path $r_u^*(t)$ representing the separation between the two zones.}}
\label{fig.case2}
\end{figure}

\begin{figure}[H]
\begin{center}
\includegraphics[width=14cm]{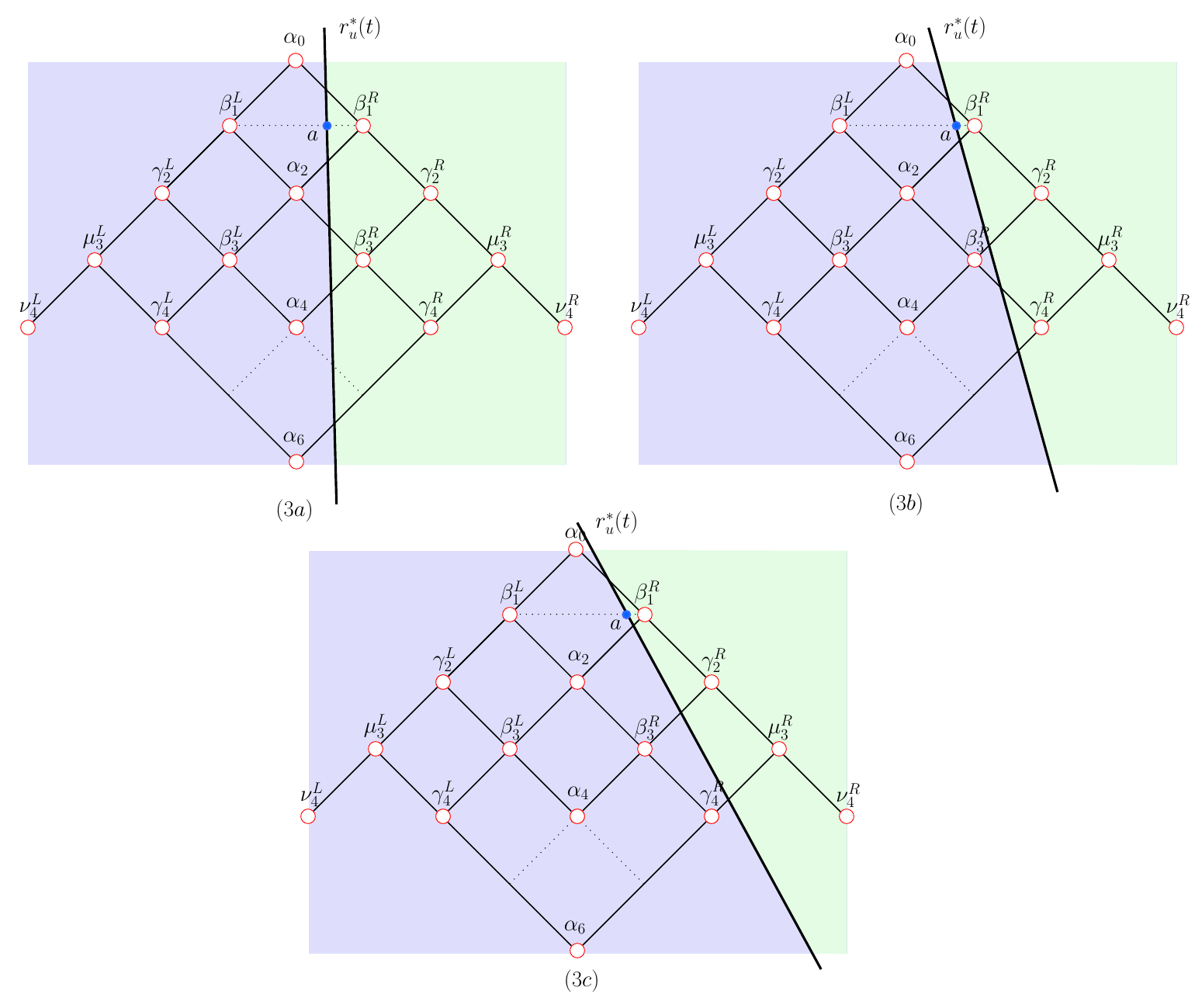}
\end{center}
\caption{\small{The three sub-cases for which the particle enters through the $[\alpha_2\beta^R_1]$ side and leaves through the $[\alpha_0\beta^R_1]$ side. The elimination of the $\Psi^-_a$ derivatives demands the utilisation of fifteen points, represented by circles, in the light cone of $\alpha_0$. Numerical efficiency suggests that the points are taken at both left and right sides of the $r_u^*(t)$ trajectory. In the three cases, the particle crosses the line $\left[\beta^L_1 \beta^R_1 \right]$ at the point 
$a$. 
The background distinguishes two zones: one where  $\Psi(r^*\!<\!r_u^*(t),t)=\Psi^-(r^*,t)$, the other where $\Psi(r^*\!>\!r_u^*(t),t)=\Psi^+(r^*,t)$, the path $r_u^*(t)$ representing the separation between the two zones.}}
\label{fig.case3}
\end{figure}

For a given cell, our aim remains the determination of the wave function value at the upper node, now rebaptised $\alpha_0$. 
As in the previous section, we consider (fifteen) points both located in the past light cone of the $\alpha_0$ point
and lying around a chosen point on the discontinuity $r_u(t)$, with the intent of determining $\Psi_{\alpha_0}$ by their linear combination. The non-regularity of the wave function due to the discontinuity, obviously entails a different value according to whether the discontinuity is approached from below ($\Psi^-$, left of the trajectory, Figs. \ref{fig.case1} - \ref{fig.case3}) or above ($\Psi^+$, right of the trajectory, Figs. \ref{fig.case1} - \ref{fig.case3}) the particle in radial fall. The same stands for the wave function derivatives. 
The addition of the jump condition to the value of the e.g. $\Psi^-$ ($\Psi^+$) wave function (or derivative of) allows to equate this sum to the value $\Psi^+$ ($\Psi^-$) of the wave function (or derivative of). This straightforward property turns being helpful for the achievement of the just mentioned linear combination of fifteen points. Incidentally, other linear combinations may be envisaged, though combinations of points located solely on one side of the discontinuity are to be avoided. 

With reference to Figs. \ref{fig.case1} - \ref{fig.case3}, there are three different cases depending upon how the trajectory of the particle crosses the cell wherein $\alpha_0$ lies. These three cases are further subdivided into three sub-cases, for
a total of nine. In the following, we label by R the points on the right of the $[\alpha_0\alpha_6]$ line and by L the points on the left. Dealing with radial fall, and thereby with a 2D code, the up and down labels might be proper; nevertheless, we stick to right and left labels, given the orientation of the $r^*$ axis in the Figs. \ref{fig.case1} - \ref{fig.case3}.    
For the first group of three, the trajectory crosses the $[\alpha_2\beta^R_1]$ and $[\alpha_0\beta^L_1]$ lines, Fig. \ref{fig.case1}; for the second group, the $[\alpha_2\beta^L_1]$ and $[\alpha_0\beta^L_1]$ lines, Fig. \ref{fig.case2};
finally for the third group, the $[\alpha_2\beta^R_1]$ and $[\alpha_0\beta^R_1]$ lines, Fig. \ref{fig.case3}. 

We start considering the sub-case (1a) shown by Fig. \ref{fig.case1}, for which the trajectory crosses the line [$\alpha_0\alpha_2$] at the point $b$. For compactness of the presentation of the final results, while we still adopt the same notation for the jump conditions, namely $[\Psi]_{q}$ for the difference $\left (\Psi^+ - \Psi^- \right )_{r_u=r_u(t_q)}$, for the jump derivatives instead, we rely henceforth on the notation $[\partial^n_{r^*}\partial^m_{t}\Psi]_q=\left (\partial^n_{r^*}\partial^m_{t}\Psi^+ - \partial^n_{r^*}\partial^m_{t}\Psi^- \right )_{r_u=r_u(t_q)}$, where $t_q$ is the coordinate time at the point $q=a,b$. We also define the lapse $\epsilon_b=t_{\alpha_0}\!-\!t_b$. 

We recall that our aim is the determination of the value of $\Psi^+_{\alpha_0}$, knowing: 
i) $\epsilon_b$, ii) the jump (analytical) conditions on $\Psi$ and its derivatives at the point $b$; iii) the values of $\Psi$ on a set of fifteen points $\left\{\alpha,\beta,\gamma,\mu,\nu\right\}$ at the left and right sides of the world line. 
A Taylor series is applied at each point around $b$ up to fourth order, thereby obtaining

\bea
\fl&\Psi^+_{\alpha_0}\!=\!\Psi^+\left(t_b\!+\!\epsilon_b,r_b^*\right)=\!\!\sum_{n=0}^4
\frac{\epsilon_b^n}{n!}\partial^{n}_t\Psi^+_b+ {\cal O}\left(\epsilon_b^5\right)~,\label{devel.1}\\
\fl&\Psi^-_{\alpha_i}\!=\!\Psi^-\left(t_b-(ih-\epsilon_b),r_b^*\right)=\!\!\sum_{n=0}^4(-1)^n\frac{(ih-\epsilon_b)^n}{n!}\partial_{t}^{n}\Psi^-_b+{\cal O}\left(h^5\right)~,\label{devel.2}\\
\fl&\Psi^\pm_{\beta^{R,L}_j}\!=\!\Psi^\pm\left(t_b\!-\!(jh\!-\!\epsilon_b),r_b^*\!\pm\!h\right)=\!\!\sum_{n+m\leq4}(-1)^{m}(\pm 1)^n\frac{h^n}{n!}\frac{(jh\!-\!\epsilon_b)^m}{m!}\partial_{r^*}^{n}\partial_{t}^{m}\Psi^\pm_b\!+\!{\cal O}\left(h^5\right)~,\label{devel.3}\nonumber\\
\fl&\\
\fl&\Psi^\pm_{\gamma^{R,L}_k}\!=\!\Psi^\pm\left(t_b\!-\!(kh\!-\!\epsilon_b),r_b^*\!\pm\!2h\right)=\!\!\sum_{n+m\leq4}(-1)^{m}(\pm 1)^n\frac{(2h)^n}{n!}\frac{(kh\!-\!\epsilon_b)^m}{m!}\partial_{r^*}^{n}\partial_{t}^{m}\Psi^\pm_b\!+\!{\cal O}\left(h^5\right)~,\label{devel.4}\nonumber\\
\fl&\\
\fl&\Psi^\pm_{\mu^{R,L}_3}\!=\!\Psi^\pm\left(t_b\!-\!(3h\!-\!\epsilon_b),r_b^*\!\pm\!3h\right)=\!\!\sum_{n+m\leq4}(-1)^{m}(\pm 1)^n\frac{(3h)^n}{n!}\frac{(3h\!-\!\epsilon_b)^m}{m!}\partial_{r^*}^{n}\partial_{t}^{m}\Psi^\pm_b\!+\!{\cal O}\left(h^5\right)~,\label{devel.5}\nonumber\\
\fl&\\
\fl&\Psi^\pm_{\nu^{R,L}_4}\!=\!\Psi^\pm\left(t_b\!-\!(4h\!-\!\epsilon_b),r_b^*\!\pm\!4h\right)=\!\!\sum_{n+m\leq4}(-1)^{m}(\pm 1)^n\frac{(4h)^n}{n!}\frac{(4h\!-\!\epsilon_b)^m}{m!}\partial_{r^*}^{n}\partial_{t}^{m}\Psi^\pm_b\!+\!{\cal O}\left(h^5\right)~,\label{devel.6}\nonumber\\
\fl&
\eea
\noindent for the indexes running as $i = 2,4,6$, $j=1,3$ and $k = 2,4$ and concerning the $\alpha$, $\beta$ and $\gamma$ nodes, respectively. Our notation implies that the subscript $R,L$ stands for $R$ when the superscript $\pm$ corresponds to $+$, whereas $R,L$ stands for $L$ when $\pm$ corresponds to $-$.
With reference to Eq. \ref{devel.1}, we get

\bea
\fl\Psi^+_{\alpha_0}&=\sum_{n=0}^4c_n\partial^{n}_t\Psi^+_b+ {\cal O}\left(h^5\right) =\sum_{n=0}^4c_n\left(\partial^{n}_t\Psi^-_b+\left[\partial^n_t\Psi\right]_b\right)+ {\cal O}\left(h^5\right) = \nonumber\\
\fl&=c_0\Psi^-_b+c_1\partial_t\Psi^-_b+c_2\partial^2_t\Psi^-_b+c_3\partial^3_t\Psi^-_b+c_4\partial^4_t\Psi^-_b+\sum_{n=0}^4c_n\left[\partial^n_t\Psi\right]_b+ {\cal O}\left(h^5\right)
\label{scheme1_with_der}
\eea

For an accuracy at fourth order, all quantities ${\cal O}(h^5)$ are disregarded. 
The sum $ {\hat S} = c_0\Psi^-_b+c_1\partial_t\Psi^-_b + c_2\partial^2_t\Psi^-_b + c_3\partial^3_t\Psi^-_b + c_4\partial^4_t\Psi^-_b$, Eq. \ref{scheme1_with_der}, is composed by numerical derivatives of lower order than ${\cal O}(h^5)$, and therefore they can't be neglected. However, the computation of high order derivatives is often accompanied by numerical noise. Therefore, we replace this sum by a combination of wave function values in the $\alpha_0$ light cone. 
This is attained in two steps. The former involves taking fifteen wave function values on the two sides of the trajectory, that is $\left\{\Psi^-_{\alpha_i}, \Psi^-_{\beta^L_j}, \Psi^+_{\beta^R_j}, \Psi^-_{\gamma^L_k}, \Psi^+_{\gamma^R_k}, \Psi^-_{\mu^L_3}, 
\Psi^+_{\mu^R_3}, \Psi^-_{\nu^L_4}, \Psi^+_{\nu^R_4} \right\}$, Fig.  
\ref{fig.case1}. The latter employs the jump conditions to relate the fifteen mentioned points with 
$\left\{\Psi^-_{\alpha_i}, \Psi^-_{\beta^L_j}, \Psi^-_{\beta^R_j}, \Psi^-_{\gamma^L_k}, \Psi^-_{\gamma^R_k}, \Psi^-_{\mu^L_3}, 
\Psi^-_{\mu^R_3}, \Psi^-_{\nu^L_4}, \Psi^-_{\nu^R_4} \right\}$. For the former step, we define the sum $S$ 

\bea 
\fl\qquad S=&
\sum_i\left(\mathcal{A}_i
\Psi^-_{\alpha_i}\right)+
\sum_j\left(\mathcal{B}^L_j
\Psi^-_{\beta^L_j}
+ \mathcal{B}^R_j
\Psi^+_{\beta^R_j}
\right)+\sum_k\left(\mathcal{G}^L_k
\Psi^-_{\gamma^L_k}
+\mathcal{G}^R_k
\Psi^+_{\gamma^R_k}
\right)\nonumber\\
\fl\qquad&+\mathcal{M}^L_3
\Psi^-_{\mu^L_3}
+\mathcal{M}^R_3
\Psi^+_{\mu^R_3}
+\mathcal{N}^L_4
\Psi^-_{\nu^L_4}
+\mathcal{N}^R_4
\Psi^+_{\nu^R_4}~.
\label{S}
\eea
where $\left\{\mathcal{A}_i,\mathcal{B}^L_j,\mathcal{B}^R_j,\mathcal{G}^L_k,\mathcal{G}^R_k,\mathcal{M}^L_3,\mathcal{M}^R_3,\mathcal{N}^L_4,\mathcal{N}^R_4\right\}$ are constants.

We observe that the $\hat S$ sum entails only wave function values at the left of the $b$ point on the trajectory. The jump conditions are once more exploited to relate the two domains $r^*<r_u^*(t)$ and $r^*>r_u^*(t)$. This specifically concerns six points 
$\left\{\beta^R_j,\gamma^R_k,\mu^R_3,\nu^R_4\right\}$. For instance, at the $\beta^R_j$ point, we can write

\bea
\fl\qquad\Psi^+_{\beta^{R}_j}&=\sum_{n+m\leq4}(-1)^{m}\frac{h^n}{n!}\frac{(jh\!-\!\epsilon_b)^m}{m!}\left(\partial_{r^*}^{n}\partial_{t}^{m}\Psi^-_b+\left[\partial^n_{r^*}\partial^m_t\Psi\right]_b\right)\!+\!{\cal O}\left(h^5\right)\nonumber\\
\fl\qquad&=\Psi^-_{\beta^{R}_j}+\sum_{n+m\leq4}(-1)^{m}\frac{h^n}{n!}\frac{(jh\!-\!\epsilon_b)^m}{m!}\left[\partial^n_{r^*}\partial^m_t\Psi\right]_b~,
\label{transfo_P+P-}
\eea 

where 
\begin{equation}
\Psi^-_{\beta^{R}_j} = \sum_{n+m\leq4}(-1)^{m}\frac{h^n}{n!}\frac{(jh\!-\!\epsilon_b)^m}{m!}\left(\partial_{r^*}^{n}\partial_{t}^{m}\Psi^-_b\right)\!+\!{\cal O}\left(h^5\right)~.
\label{notation_P-}
\end{equation}

By application of the same transformation to the quantities $\Psi^+_{\gamma^R_k},\Psi^+_{\mu^R_3},\Psi^+_{\nu^R_3}$, Eq. \ref{S} becomes

\bea \fl S - \Phi^{jump}_{r_u^*}=&\sum_i\left(\mathcal{A}_i\Psi^-_{\alpha_i}\right)+\sum_j\left(\mathcal{B}^L_j\Psi^-_{\beta^L_j}+\mathcal{B}^R_j\Psi^-_{\beta^R_j}\right)+\sum_k\left(\mathcal{G}^L_k\Psi^-_{\gamma^L_k}+\mathcal{G}^R_k\Psi^-_{\gamma^R_k}\right)\nonumber\\
\fl&+\mathcal{M}^L_3\Psi^-_{\mu^L_3}+\mathcal{M}^R_3\Psi^-_{\mu^R_3}+\mathcal{N}^L_4\Psi^-_{\nu^L_4}+\mathcal{N}^R_4\Psi^-_{\nu^R_4}~,
\label{s-phi}
\eea
where $\Phi^{jump}_{r_u^*}$ is an analytic function, composed by the jump conditions at the $b$ point, weighted by  coefficients issued by Eq. \ref{transfo_P+P-} or similar equations.

Having only $\Psi^-$ terms on the right hand side of Eq. \ref{s-phi}, we can finally search the coefficients $\left\{\mathcal{A}_i,\mathcal{B}^L_j,\mathcal{B}^R_j,\mathcal{G}^L_k,\mathcal{G}^R_k,\mathcal{M}^L_3,\mathcal{M}^R_3,\mathcal{N}^L_4,\mathcal{N}^R_4\right\}$ that satisfy the equation ${\hat S} = S - \Phi^{jump}_{r_u^*}$, that is  

\bea
\fl\qquad&c_0\Psi^-_b+c_1\partial_t\Psi^-_b + c_2\partial^2_t\Psi^-_b + c_3\partial^3_t\Psi^-_b + c_4\partial^4_t\Psi^-_b =\nonumber\\ \fl\qquad&\sum_i\left(\mathcal{A}_i\Psi^-_{\alpha_i}\right)+\sum_j\left(\mathcal{B}^L_j\Psi^-_{\beta^L_j}+\mathcal{B}^R_j\Psi^-_{\beta^R_j}\right)+\sum_k\left(\mathcal{G}^L_k\Psi^-_{\gamma^L_k}+\mathcal{G}^R_k\Psi^-_{\gamma^R_k}\right)\nonumber\\
\fl\qquad&+\mathcal{M}^L_3\Psi^-_{\mu^L_3}+\mathcal{M}^R_3\Psi^-_{\mu^R_3}+\mathcal{N}^L_4\Psi^-_{\nu^L_4}+\mathcal{N}^R_4\Psi^-_{\nu^R_4}~.
\label{syst_explicit}
\eea

Using the notation of Eqs. \ref{transfo_P+P-}, \ref{notation_P-}, and by injection of Eqs. \ref{devel.2}-\ref{devel.6}, 
a Taylor expansion of fourth order at the $b$ point is applied
to the right-hand side of Eq. \ref{syst_explicit}. The system can be cast in a matrix form

{
\beq
\qquad\qquad\qquad\qquad\mathbb{T}\cdot\mathbb{P}=\mathbb{C}~,
\eeq
}

\noindent { where $\mathbb{P}$} is the unknown 15-vector formed by the coefficients $\left\{\mathcal{A}_i,\mathcal{B}^L_j,\mathcal{B}^R_j,\mathcal{G}^L_k,\mathcal{G}^R_k,\mathcal{M}^L_3,\right.$\\$\left.\mathcal{M}^R_3,\mathcal{N}^L_4,\mathcal{N}^R_4\right\}$

{
\beq
\fl\qquad\mathbb{P}=\left(\mathcal{A}_2,\mathcal{A}_4,\mathcal{A}_6,\mathcal{B}^L_1,\mathcal{B}^L_3,\mathcal{B}^R_1,\mathcal{B}^R_3,\mathcal{G}^L_2,\mathcal{G}^L_4,\mathcal{G}^R_2,\mathcal{G}^R_4,\mathcal{M}^L_3,\mathcal{M}^R_3,\mathcal{N}^L_4,\mathcal{N}^R_4\right)^t~,
\eeq

\noindent and $\mathbb{C}$ is given by the 15-vector
\beq
\qquad\mathbb{C}=(c_0,\ c_1,\ c_2,\ c_3,\ c_4,\ 0,\cdots,\ 0)^t~,
\eeq
}

\noindent while $\mathbb{T}$ is the $(15\times 15)$ matrix constructed from the Taylor coefficients in Eqs. \ref{devel.2}-\ref{devel.6} { (see appendix). By inversion of $\mathbb{T}$, we get $\mathbb{P}$ and specifically}

\bea
\qquad  \mathcal{A}_2 = \frac{-27}{5}~,\ 
\quad  \mathcal{A}_4 = \frac{-9}5~,\ 
\quad  \mathcal{A}_6 = \frac{1}{5}~,\nonumber\\
\qquad  \mathcal{B}^L_1 = \mathcal{B}^R_1 = \frac{12}5~,\quad
\qquad  \mathcal{B}^L_3 = \mathcal{B}^R_3 = \frac{18}5~,\nonumber\\
\qquad  \mathcal{G}^L_2 = \mathcal{G}^R_2 = \frac{-9}5~,\quad
\qquad  \mathcal{G}^L_4 = \mathcal{G}^R_4 = \frac{-3}5~,\nonumber\\
\qquad  \mathcal{M}^L_3 = \mathcal{M}^R_3 = \frac{2}5~,\quad
\qquad  \mathcal{N}^L_4 = \mathcal{N}^R_4 = 0~.\nonumber
\eea

The following equivalences path the last stretch of the way
\beq
{\Psi^+_{\alpha_0}} = S-\Phi^{jump}_{r_u^*}+\sum_{n=0}^4c_n\left[\partial^n_t\Psi\right]_b = S  + 
\Phi^{(1)}_{r_u^*(t_b)}~,
\eeq

\nid and explicitly, we get  

\bea
\fl{\Psi^+_{\alpha_0}}=&
-\frac{27}{5}\Psi^-_{\alpha_2} - 
\frac{9}{5}\Psi^-_{\alpha_4} +
\frac{1}{5}\Psi^-_{\alpha_6} +
\frac{12}{5}\left(\Psi^-_{\beta^L_1}+\Psi^+_{\beta^R_1}\right) + 
\frac{18}{5}\left(\Psi^-_{\beta^L_3}+\Psi^\pm_{\beta^R_3}\right) \nonumber\\
\fl&-\frac{9}{5}\left(\Psi^-_{\gamma^L_2}+\Psi^+_{\gamma^R_2}\right) + 
\frac{3}{5}\left(\Psi^-_{\gamma^L_4}+\Psi^\pm_{\gamma^R_4}\right) -
\frac{2}{5}\left(\Psi^-_{\mu^L_3}+\Psi^+_{\mu^R_3}\right)+\Phi^{(1)}_{r_u^*(t_b)}~, 
\label{case1final}
\eea

\nid where $\Psi^\pm_{\beta^R_3} = \Psi^+_{\beta^R_3}$ for sub-case (1a), and 
$\Psi^\pm_{\beta^R_3} = \Psi^-_{\beta^R_3}$ for sub-cases (1b,1c); $\Psi^\pm_{\gamma^R_4} = \Psi^+_{\gamma^R_4}$ for sub-cases (1a,1b), and $\Psi^\pm_{\gamma^R_4} = \Psi^-_{\gamma^R_4}$ for sub-case (1c); and  $\Phi^{(1)}_{r_u^*(t_b)}$ is an analytic function, that for the (1a) sub-case, it takes the value 

\bea
\fl\Phi^{(1a)}_{r_u^*(t_b)} &= -3\left[\Psi\right]_b
-\frac{3\,\left( 5\,\epsilon_b-14\,h\right) }{5}\left[\partial_t\Psi\right]_b 
-\frac{3\,\left( \epsilon_b-2\,h\right) \,\left( 5\,\epsilon_b-18\,h\right) }{10}\left[\partial^2_t\Psi\right]_b\nonumber\\
\fl&-\frac{5\,{\epsilon_b}^{3}-42\,h\,{\epsilon_b}^{2}+108\,{h}^{2}\,\epsilon_b-96\,{h}^{3}}{10}\left[\partial^3_t\Psi\right]_b\nonumber\\
\fl&-\frac{5\,{\epsilon_b}^{4}-56\,h\,{\epsilon_b}^{3}+216\,{h}^{2}\,{\epsilon_b}^{2}-384\,{h}^{3}\,\epsilon_b+240\,{h}^{4}}{40}\left[\partial^4_t\Psi\right]_b -\frac{12\,h}{5}\left[\partial_{r^*}\Psi\right]_b\nonumber\\
\fl& +\frac{2\,{h}^{3}}{5}\left[\partial^3_{r^*}\Psi\right]_b
-\frac{12\,h\,\left( \epsilon_b-2\,h\right) }{5}\left[\partial_{r^*}\partial_t\Psi\right]_b 
-\frac{6\,h\,\left( {\epsilon_b}^{2}-4\,h\,\epsilon_b+5\,{h}^{2}\right) }{5}\left[\partial^3_{r^*}\partial_t\Psi\right]_b\nonumber\\
\fl& + \frac{2\,{h}^{3}\,\left( \epsilon_b-h\right) }{5}\left[\partial^2_{r^*}\partial^2_t\Psi\right]_b
-\frac{2\,h\,\left( {\epsilon_b}^{3}-6\,h\,{\epsilon_b}^{2}+15\,{h}^{2}\,\epsilon_b-11\,{h}^{3}\right) }{5}\left[\partial_{r^*}\partial^3_t\Psi\right]_b
~.\nonumber\\
\fl&\ 
\eea

The quantity $\Phi^{(1)}_{r_u^*}$ varies according to the different sub-cases: for the case (1b) of Fig. \ref{fig.case1}, the point  
$\beta^R_3$, whereas for the case (1c) the points $\beta^R_3$ and $\gamma^R_4$ are in the $r^*<r_u^*$ domain. 
Therefore 

\bea
\fl\Phi^{(1b)}_{r_u^*(t_b)} &= 
\frac{6}{10}\left[\Psi\right]_b
+\frac{3\,\left( \epsilon_b-4\,h\right) }{5}\left[\partial_t\Psi\right]_b
+\frac{3\,\left( {\epsilon_b}^{2}-8\,h\,\epsilon_b+18\,{h}^{2}\right) }{10}\left[\partial^2_t\Psi\right]_b\nonumber\\
\fl&+\frac{{\epsilon_b}^{3}-12\,h\,{\epsilon_b}^{2}+54\,{h}^{2}\,\epsilon_b-66\,{h}^{3}}{10}\left[\partial^3_t\Psi\right]_b\nonumber\\
\fl&+\frac{{\epsilon_b}^{4}-16\,h\,{\epsilon_b}^{3}+108\,{h}^{2}\,{\epsilon_b}^{2}-264\,{h}^{3}\,\epsilon_b+246\,{h}^{4}}{40}\left[\partial^4_t\Psi\right]_b\nonumber\\
\fl&+\frac{6\,h}{5}\left[\partial_{r^*}\Psi\right]_b
+\frac{9\,{h}^{2}}{5}\left[\partial^2_{r^*}\Psi\right]_b
+{h}^{3}\left[\partial^3_{r^*}\Psi\right]_b
+\frac{3\,{h}^{4}}{20}\left[\partial^4_{r^*}\Psi\right]_b\nonumber\\
\fl&+\frac{6\,h\,\left( \epsilon_b-5\,h\right) }{5}\left[\partial_{r^*}\partial_t\Psi\right]_b
+\frac{9\,{h}^{2}\,\left( \epsilon_b-3\,h\right) }{5}\left[\partial^2_{r^*}\partial_t\Psi\right]_b\nonumber\\
\fl&+\frac{3\,h\,\left( {\epsilon_b}^{2}-10\,h\,\epsilon_b+17\,{h}^{2}\right) }{5}\left[\partial^3_{r^*}\partial_t\Psi\right]_b
+\frac{9\,{h}^{2}\,{\left( \epsilon_b-3\,h\right) }^{2}}{10}\left[\partial_{r^*}\partial^2_t\Psi\right]_b\nonumber\\
\fl&
+\frac{{h}^{3}\,\left( 5\,\epsilon_b-11\,h\right) }{5}\left[\partial^2_{r^*}\partial^2_t\Psi\right]_b
+\frac{h\,\left( {\epsilon_b}^{3}-15\,h\,{\epsilon_b}^{2}+51\,{h}^{2}\,\epsilon_b-59\,{h}^{3}\right) }{5}\left[\partial_{r^*}\partial^3_t\Psi\right]_b~,\nonumber\\
\fl&
\eea

\bea
\fl\Phi^{(1c)}_{r_u^*(t_b)} &=
\frac{3\,{h}^{2}}{5}\left[\partial^2_t\Psi\right]_b
+\frac{{h}^{2}\,\left( 3\,\epsilon_b-h\right) }{5}\left[\partial^3_t\Psi\right]_b
+\frac{{h}^{2}\,\left( 6\,{\epsilon_b}^{2}-4\,h\,\epsilon_b-5\,{h}^{2}\right) }{20}\left[\partial^4_t\Psi\right]_b\nonumber\\
\fl&+\frac{3\,{h}^{2}}{5}\left[\partial^2_{r^*}\Psi\right]_b
+\frac{{h}^{3}}{5}\left[\partial^3_{r^*}\Psi\right]_b
-\frac{{h}^{4}}{4}\left[\partial^4_{r^*}\Psi\right]_b
-\frac{6\,{h}^{2}}{5}\left[\partial_{r^*}\partial_t\Psi\right]_b\nonumber\\
\fl&+\frac{3\,{h}^{2}\,\left( \epsilon_b-h\right) }{5}\left[\partial^2_{r^*}\partial_t\Psi\right]_b
-\frac{3\,{h}^{2}\,\left( 2\,\epsilon_b-h\right) }{5}\left[\partial^3_{r^*}\partial_t\Psi\right]_b\nonumber\\
\fl&+\frac{3\,{h}^{2}\,\left( {\epsilon_b}^{2}-2\,h\,\epsilon_b-5\,{h}^{2}\right) }{10}\left[\partial_{r^*}\partial^2_t\Psi\right]_b  +\frac{{h}^{3}\,\left( \epsilon_b+5\,h\right) }{5}\left[\partial^2_{r^*}\partial^2_t\Psi\right]_b\nonumber\\
\fl&
-\frac{{h}^{2}\,\left( 3\,{\epsilon_b}^{2}-3\,h\,\epsilon_b-5\,{h}^{2}\right) }{5}\left[\partial_{r^*}\partial^3_t\Psi\right]_b
~.
\eea

We thus have obtained, without direct integration of the singular source and the potential term, the value of the upper node. The equations shows three types of terms: the preceding node values of the same cell, the jump conditions which are fully analytical quantities, and the wave function values at adjacent cells. Incidentally, at first order \cite{aosp11b}, the latter type of terms disappears and a simpler expression is obtained. 

Similar relations are found for the other two remaining cases. For case 2, Fig. \ref{fig.case2}, we obtain (having defined the shift $\epsilon_a=t_{\beta^R_1}\!-\!r^*_a$)

\bea
\fl{\Psi^+_{\alpha_0}}=&
-\frac{27}{5}\Psi^+_{\alpha_2} - 
\frac{9}{5}\Psi^\pm_{\alpha_4} +
\frac{1}{5}\Psi^\pm_{\alpha_6} +
\frac{12}{5}\left(\Psi^-_{\beta^L_1}+\Psi^+_{\beta^R_1}\right) + 
\frac{18}{5}\left(\Psi^-_{\beta^L_3}+\Psi^+_{\beta^R_3}\right) \nonumber\\
\fl&-\frac{9}{5}\left(\Psi^-_{\gamma^L_2}+\Psi^+_{\gamma^R_2}\right) + 
\frac{3}{5}\left(\Psi^-_{\gamma^L_4}+\Psi^+_{\gamma^R_4}\right) -
\frac{2}{5}\left(\Psi^-_{\mu^L_3}+\Psi^+_{\mu^R_3}\right)+\Phi^{(2)}_{r_u^*(t_a)}~. 
\eea

\nid where $\Psi^\pm_{\alpha_4} = \Psi^+_{\alpha_4}$ for sub-cases (2a,2b), and 
$\Psi^\pm_{\alpha_4} = \Psi^-_{\alpha_4}$ for sub-case (2c); $\Psi^\pm_{\alpha_6} = \Psi^+_{\alpha_6}$ for sub-case (2a), and 
$\Psi^\pm_{\alpha_6} = \Psi^-_{\alpha_6}$ for sub-cases (2b,2c). For the (2a) sub-case, $\Phi^{(2)}_{r_u^*(t_a)}$ takes the following value

\bea
\fl\Phi^{(2a)}_{r_u^*(t_a)}&=4\left[\Psi\right]_a
-\frac{22\,h}{5}\left[\partial_t\Psi\right]_a
+\frac{22\,{h}^{2}}{5}\left[\partial^2_t\Psi\right]_a
-\frac{7\,{h}^{3}}{3}\left[\partial^3_t\Psi\right]_a
+\frac{17\,{h}^{4}}{30}\left[\partial^4_t\Psi\right]_a\nonumber\\
\fl&+\frac{4\,\left( 5\,\epsilon_a-8\,h\right) }{5}\left[\partial_{r^*}\Psi\right]_a
+\frac{2\,\left( \epsilon_a-h\right) \,\left( 5\,\epsilon_a-11\,h\right) }{5}\left[\partial^2_{r^*}\Psi\right]_a\nonumber\\
\fl&+\frac{2\,\left( 5\,{\epsilon_a}^{3}-24\,h\,{\epsilon_a}^{2}+33\,{h}^{2}\,\epsilon_a-11\,{h}^{3}\right) }{15}\left[\partial^3_{r^*}\Psi\right]_a\nonumber\\
\fl&+\frac{\left( \epsilon_a-h\right) \,\left( 5\,{\epsilon_a}^{3}-27\,h\,{\epsilon_a}^{2}+39\,{h}^{2}\,\epsilon_a-5\,{h}^{3}\right) }{30}\left[\partial^4_{r^*}\Psi\right]_a\nonumber\\
\fl&-\frac{2\,h\,\left( 11\,\epsilon_a-17\,h\right) }{5}\left[\partial_{r^*}\partial_t\Psi\right]_a
-\frac{h\,\left( \epsilon_a-h\right) \,\left( 11\,\epsilon_a-23\,h\right) }{5}\left[\partial^2_{r^*}\partial_t\Psi\right]_a\nonumber\\
\fl&
+\frac{2\,{h}^{2}\,\left( 11\,\epsilon_a-17\,h\right) }{5}\left[\partial_{r^*}\partial^2_t\Psi\right]_a
-\frac{h\,{\left( \epsilon_a-h\right) }^{2}\,\left( 11\,\epsilon_a-29\,h\right) }{15}\left[\partial^3_{r^*}\partial_t\Psi\right]_a
\nonumber\\
\fl&+\frac{{h}^{3}\,\left( \epsilon_b+5\,h\right) }{5}\left[\partial^2_{r^*}\partial^2_t\Psi\right]_b
-\frac{{h}^{3}\,\left( 35\,\epsilon_a-41\,h\right) }{15}\left[\partial_{r^*}\partial^3_t\Psi\right]_a
~.
\eea

For the same preceding reason, the sub-cases (2b, 2c) differ as the points $\alpha_4$ and $\alpha_6$ are or aren't in the 
$r^*>r_u^*$ domain. Therefore, we have

\bea
\fl\Phi^{(2b)}_{r_u^*(t_a)}&=
\frac{42}{10}\left[\Psi\right]_a
-\frac{27\,h}{5}\left[\partial_t\Psi\right]_a
+\frac{69\,{h}^{2}}{10}\left[\partial^2_t\Psi\right]_a
-\frac{13\,{h}^{3}}{2}\left[\partial^3_t\Psi\right]_a
+\frac{231\,{h}^{4}}{40}\left[\partial^4_t\Psi\right]_a\nonumber\\
\fl&
+\frac{3\,\left( 7\,\epsilon_a-11\,h\right) }{5}\left[\partial_{r^*}\Psi\right]_a
+\frac{3\,\left( \epsilon_a-h\right) \,\left( 7\,\epsilon_a-15\,h\right) }{10}\left[\partial^2_{r^*}\Psi\right]_a\nonumber\\
\fl&
+\frac{7\,{\epsilon_a}^{3}-33\,h\,{\epsilon_a}^{2}+45\,{h}^{2}\,\epsilon_a-15\,{h}^{3}}{10}\left[\partial^3_{r^*}\Psi\right]_a\nonumber\\
\fl&
+\frac{\left( \epsilon_a-h\right) \,\left( 7\,{\epsilon_a}^{3}-37\,h\,{\epsilon_a}^{2}+53\,{h}^{2}\,\epsilon_a-7\,{h}^{3}\right) }{40}\left[\partial^4_{r^*}\Psi\right]_a\nonumber\\
\fl&
-\frac{3\,h\,\left( 9\,\epsilon_a-13\,h\right) }{5}\left[\partial_{r^*}\partial_t\Psi\right]_a
-\frac{3\,h\,\left( \epsilon_a-h\right) \,\left( 9\,\epsilon_a-17\,h\right) }{10}\left[\partial^2_{r^*}\partial_t\Psi\right]_a
\nonumber\\
\fl&
+\frac{3\,{h}^{2}\,\left( 23\,\epsilon_a-31\,h\right) }{10}\left[\partial_{r^*}\partial^2_t\Psi\right]_a
-\frac{3\,h\,{\left( \epsilon_a-h\right) }^{2}\,\left( 3\,\epsilon_a-7\,h\right) }{10}\left[\partial^3_{r^*}\partial_t\Psi\right]_a \nonumber\\
\fl&
+\frac{{h}^{2}\,\left( \epsilon_a-h\right) \,\left( 157\,\epsilon_a-109\,h\right) }{20}\left[\partial^2_{r^*}\partial^2_t\Psi\right]_a
-\frac{{h}^{3}\,\left( 65\,\epsilon_a-69\,h\right) }{10}\left[\partial_{r^*}\partial^3_t\Psi\right]_a
~, \nonumber\\
\fl&\
\eea

\bea
\fl\Phi^{(2c)}_{r_u^*(t_a)}&=
\frac{24}{10}\left[\Psi\right]_a
-\frac{6\,{h}^{2}}{5}\left[\partial^2_t\Psi\right]_a
+\frac{8\,{h}^{3}}{5}\left[\partial^3_t\Psi\right]_a
-\frac{3\,{h}^{4}}{10}\left[\partial^4_t\Psi\right]_a\nonumber\\
\fl&
+\frac{12\,\left( \epsilon_a-2\,h\right) }{5}\left[\partial_{r^*}\Psi\right]_a
+\frac{6\,\left( \epsilon_a-3\,h\right) \,\left( \epsilon_a-h\right) }{5}\left[\partial^2_{r^*}\Psi\right]_a\nonumber\\
\fl&
+\frac{2\,\left( {\epsilon_a}^{3}-6\,h\,{\epsilon_a}^{2}+9\,{h}^{2}\,\epsilon_a-3\,{h}^{3}\right) }{5}\left[\partial^3_{r^*}\Psi\right]_a\nonumber\\
\fl&
+\frac{\left( \epsilon_a-h\right) \,\left( {\epsilon_a}^{3}-7\,h\,{\epsilon_a}^{2}+11\,{h}^{2}\,\epsilon_a-{h}^{3}\right) }{10}\left[\partial^4_{r^*}\Psi\right]_a 
+\frac{12\,{h}^{2}}{5}\left[\partial_{r^*}\partial_t\Psi\right]_a
\nonumber\\
\fl&
+\frac{12\,{h}^{2}\,\left( \epsilon_a-h\right) }{5}\left[\partial^2_{r^*}\partial_t\Psi\right]_a
-\frac{6\,{h}^{2}\,\left( \epsilon_a+h\right) }{5}\left[\partial_{r^*}\partial^2_t\Psi\right]_a \nonumber \\
\fl&
+\frac{6\,{h}^{2}\,{\left( \epsilon_a-h\right) }^{2}}{5}\left[\partial^3_{r^*}\partial_t\Psi\right]_a
+\frac{{h}^{2}\,\left( \epsilon_a-h\right) \,\left( 19\,\epsilon_a-7\,h\right) }{5}\left[\partial^2_{r^*}\partial^2_t\Psi\right]_a
\nonumber\\
\fl&
+\frac{2\,{h}^{3}\,\left( 4\,\epsilon_a-3\,h\right) }{5}\left[\partial_{r^*}\partial^3_t\Psi\right]_a
~.
\eea

Finally for case 3, Fig. \ref{fig.case3}, we have

\bea
\fl{\Psi^-_{\alpha_0}}=&
-\frac{27}{5}\Psi^-_{\alpha_2} - 
\frac{9}{5}\Psi^-_{\alpha_4} +
\frac{1}{5}\Psi^-_{\alpha_6} +
\frac{12}{5}\left(\Psi^-_{\beta^L_1}+\Psi^+_{\beta^R_1}\right) + 
\frac{18}{5}\left(\Psi^-_{\beta^L_3}+\Psi^\pm_{\beta^R_3}\right) \nonumber\\
\fl&-\frac{9}{5}\left(\Psi^-_{\gamma^L_2}+\Psi^+_{\gamma^R_2}\right) + 
\frac{3}{5}\left(\Psi^-_{\gamma^L_4}+\Psi^\pm_{\gamma^R_4}\right) -
\frac{2}{5}\left(\Psi^-_{\mu^L_3}+\Psi^+_{\mu^R_3}\right)+\Phi^{(3)}_{r_u^*(t_a)}~, 
\eea

\nid where $\Psi^\pm_{\beta^R_3} = \Psi^+_{\beta^R_3}$ for sub-case (3a), and 
$\Psi^\pm_{\beta^R_3} = \Psi^-_{\beta^R_3}$ for sub-cases (3b,3c); $\Psi^\pm_{\gamma^R_4} = \Psi^+_{\gamma^R_4}$ for sub-cases (3a,3b), and $\Psi^\pm_{\gamma^R_4} = \Psi^-_{\gamma^R_4}$ for sub-case (3c); and  $\Phi^{(3)}_{r_u^*(t_a)}$ takes the values 

\bea
\fl\Phi^{(3a)}_{r_u^*(t_a)}&=-4\left[\Psi\right]_a
+\frac{22\,h}{5}\left[\partial_t\Psi\right]_a
-\frac{22\,{h}^{2}}{5}\left[\partial^2_t\Psi\right]_a
+\frac{7\,{h}^{3}}{3}\left[\partial^3_t\Psi\right]_a
-\frac{17\,{h}^{4}}{30}\left[\partial^4_t\Psi\right]_a\nonumber\\
\fl&-\frac{4\,\left( 5\,\epsilon_a-2\,h\right) }{5}\left[\partial_{r^*}\Psi\right]_a
-\frac{2\,\left( \epsilon_a-h\right) \,\left( 5\,\epsilon_a+h\right) }{5}\left[\partial^2_{r^*}\Psi\right]_a\nonumber\\
\fl&-\frac{2\,\left( 5\,{\epsilon_a}^{3}-6\,h\,{\epsilon_a}^{2}-3\,{h}^{2}\,\epsilon_a+{h}^{3}\right) }{15}\left[\partial^3_{r^*}\Psi\right]_a\nonumber\\
\fl&-\frac{\left( \epsilon_a-h\right) \,\left( 5\,{\epsilon_a}^{3}-3\,h\,{\epsilon_a}^{2}-9\,{h}^{2}\,\epsilon_a-5\,{h}^{3}\right) }{30}\left[\partial^4_{r^*}\Psi\right]_a\nonumber\\
\fl&+\frac{2\,h\,\left( 11\,\epsilon_a-5\,h\right) }{5}\left[\partial_{r^*}\partial_t\Psi\right]_a
+\frac{h\,\left( \epsilon_a-h\right) \,\left( 11\,\epsilon_a+h\right) }{5}\left[\partial^2_{r^*}\partial_t\Psi\right]_a\nonumber\\
\fl&
-\frac{2\,{h}^{2}\,\left( 11\,\epsilon_a-5\,h\right) }{5}\left[\partial_{r^*}\partial^2_t\Psi\right]_a
+\frac{h\,{\left( \epsilon_a-h\right) }^{2}\,\left( 11\,\epsilon_a+7\,h\right) }{15}\left[\partial^3_{r^*}\partial_t\Psi\right]_a
\nonumber\\
\fl&
+\frac{{h}^{2}\,\left( \epsilon_a-h\right) \,\left( 11\,\epsilon_a+h\right) }{5}\left[\partial^2_{r^*}\partial^2_t\Psi\right]_a
+\frac{{h}^{3}\,\left( 35\,\epsilon_a-29\,h\right) }{15}\left[\partial_{r^*}\partial^3_t\Psi\right]_a
~,
\eea

\bea
\fl\Phi^{(3b)}_{r_u^*(t_a)}&=
-\frac{2}{5}\left[\Psi\right]_a
-\frac{14\,h}{5}\left[\partial_t\Psi\right]_a
+\frac{14\,{h}^{2}}{5}\left[\partial^2_t\Psi\right]_a
-\frac{37\,{h}^{3}}{15}\left[\partial^3_t\Psi\right]_a
+\frac{11\,{h}^{4}}{6}\left[\partial^4_t\Psi\right]_a\nonumber\\
\fl&
-\frac{2\,\left( \epsilon_a-4\,h\right) }{5}\left[\partial_{r^*}\Psi\right]_a
-\frac{{\epsilon_a}^{2}-8\,h\,\epsilon_a-2\,{h}^{2}}{5}\left[\partial^2_{r^*}\Psi\right]_a\nonumber\\
\fl&
-\frac{{\epsilon_a}^{3}-12\,h\,{\epsilon_a}^{2}-6\,{h}^{2}\,\epsilon_a+2\,{h}^{3}}{15}\left[\partial^3_{r^*}\Psi\right]_a\nonumber\\
\fl&
-\frac{{\epsilon_a}^{4}-16\,h\,{\epsilon_a}^{3}-12\,{h}^{2}\,{\epsilon_a}^{2}+8\,{h}^{3}\,\epsilon_a+10\,{h}^{4}}{60}\left[\partial^4_{r^*}\Psi\right]_a\nonumber\\
\fl&
-\frac{2\,h\,\left( 7\,\epsilon_a+5\,h\right) }{5}\left[\partial_{r^*}\partial_t\Psi\right]_a
-\frac{h\,\left( 7\,{\epsilon_a}^{2}+10\,h\,\epsilon_a+{h}^{2}\right) }{5}\left[\partial^2_{r^*}\partial_t\Psi\right]_a\nonumber\\
\fl&
+\frac{2\,{h}^{2}\,\left( 7\,\epsilon_a+5\,h\right) }{5}\left[\partial_{r^*}\partial^2_t\Psi\right]_a
-\frac{h\,\left( 7\,{\epsilon_a}^{3}+15\,h\,{\epsilon_a}^{2}+3\,{h}^{2}\,\epsilon_a-7\,{h}^{3}\right) }{15}\left[\partial^3_{r^*}
\partial_t\Psi\right]_a\nonumber\\
\fl&
-\frac{{h}^{2}\,\left( 7\,{\epsilon_a}^{2}+10\,h\,\epsilon_a+{h}^{2}\right) }{5}\left[\partial^2_{r^*}\partial^2_t\Psi\right]_a
-\frac{{h}^{3}\,\left( 37\,\epsilon_a+29\,h\right) }{15}\left[\partial_{r^*}\partial^3_t\Psi\right]_a~,
\eea 

\bea
\fl\Phi^{(3c)}_{r_u^*(t_a)}&=
-\left[\Psi\right]_a
-h\left[\partial_t\Psi\right]_a
+\frac{{h}^{2}}{10}\left[\partial^2_t\Psi\right]_a
+\frac{7\,{h}^{3}}{30}\left[\partial^3_t\Psi\right]_a
-\frac{23\,{h}^{4}}{120}\left[\partial^4_t\Psi\right]_a\nonumber\\
\fl&
+(h-\epsilon_a)\left[\partial_{r^*}\Psi\right]_a
-\frac{5\,{\epsilon_a}^{2}-10\,h\,\epsilon_a-{h}^{2}}{10}\left[\partial^2_{r^*}\Psi\right]_a\nonumber\\
\fl&
-\frac{5\,{\epsilon_a}^{3}-15\,h\,{\epsilon_a}^{2}-3\,{h}^{2}\,\epsilon_a+7\,{h}^{3}}{30}\left[\partial^3_{r^*}\Psi\right]_a\nonumber\\
\fl&
-\frac{5\,{\epsilon_a}^{4}-20\,h\,{\epsilon_a}^{3}-6\,{h}^{2}\,{\epsilon_a}^{2}+28\,{h}^{3}\,\epsilon_a+23\,{h}^{4}}{120}\left[\partial^4_{r^*}\Psi\right]_a\nonumber\\
\fl&
-\frac{h\,\left( 5\,\epsilon_a+h\right) }{5}\left[\partial_{r^*}\partial_t\Psi\right]_a
-\frac{h\,\left( \epsilon_a-h\right) \,\left( 5\,\epsilon_a+7\,h\right) }{10}\left[\partial^2_{r^*}\partial_t\Psi\right]_a\nonumber\\
\fl&
+\frac{{h}^{2}\,\left( \epsilon_a-7\,h\right) }{10}\left[\partial_{r^*}\partial^2_t\Psi\right]_a
-\frac{h\,\left( 5\,{\epsilon_a}^{3}+3\,h\,{\epsilon_a}^{2}-21\,{h}^{2}\,\epsilon_a-23\,{h}^{3}\right) }{30}\left[\partial^3_{r^*}\partial_t\Psi\right]_a
\nonumber\\
\fl&
-\frac{{h}^{2}\,\left( {\epsilon_a}^{2}-14\,h\,\epsilon_a-23\,{h}^{2}\right) }{20}\left[\partial^2_{r^*}\partial^2_t\Psi\right]_a
+\frac{{h}^{3}\,\left( 7\,\epsilon_a+23\,h\right) }{30}\left[\partial_{r^*}\partial^3_t\Psi\right]_a~.
\eea

\noindent The jump conditions in the tortoise $r^*$ relate to those previously computed in the $r$ variable (the relations for mixed derivatives $(r^*,t)$ are easily inferred)

\bea
\fl\left[\Psi_{,r^*}\right]&=&f_{r_u}\left[\Psi_{,r}\right]~,\\
\fl\left[\Psi_{,r^*r^*}\right]&=&f_{r_u}f'_{r_u}\left[\Psi_{,r}\right]+f^2_{r_u}\left[\Psi_{,rr}\right]~,\\
\fl\left[\Psi_{,r^*r^*r^*}\right]&=&f_{r_u}\left(f'^2+ff''\right)_{r_u}\left[\Psi_{,r}\right]+3f_{r_u}^2f'_{r_u}\left[\Psi_{,rr}\right]+f_{r_u}^3\left[\Psi_{,rrr}\right]~,\\
\fl\left[\Psi_{,r^*r^*r^*r^*}\right]&=&f_{r_u}\left(f'^3+4ff'f''+f^2f^{'''}\right)_{r_u}\left[\Psi_{,r}\right]+f_{r_u}^2\left(7f'^2+4ff''\right)_{r_u}\left[\Psi_{,rr}\right]\nonumber\\
\fl &\ &+6f_{r_u}^3f_{r_u}'\left[\Psi_{,rrr}\right]+f_{r_u}^4\left[\Psi_{,rrrr}\right]~.
\eea

{ 
\section{Numerical implementation}

Waveforms at infinity and at the particle position  at first order are to be
found in \cite{aosp11b}, as well as comparisons with other methods. Herein we are concerned on the numerical improvement. To this end, we have considered a distant observer, located at $r^*=400(2M)$. The observer is reached by a pulse produced by a Gaussian, time-symmetric perturbation

\begin{eqnarray}
\Psi(r^*,t)_{t=0}&=&\exp\left[-(r^*-r_0^*)^2\right]~,\\
\partial_t\Psi(r^*,t)_{t=0}&=&0~.
\end{eqnarray}

Fig. \ref{fig:Psi_h01}, obtained for $r_{u0}=5(2M)$, shows the waveform produced in the homogeneous case. The convergence rate is computed as ($\epsilon^{(n)}(\xi)$ is the unknown error function of order $\approx1$)
\beq
n=\log\left|\frac{\Psi(4h)-\Psi(2h)}{\Psi(2h)-\Psi(h)}\right|/\log(2)
+\log\left|\epsilon^{(n)}(\xi)\right|/\log(2)~.
\eeq

\begin{figure}[H]
\begin{center}
\includegraphics[width=8.5cm]{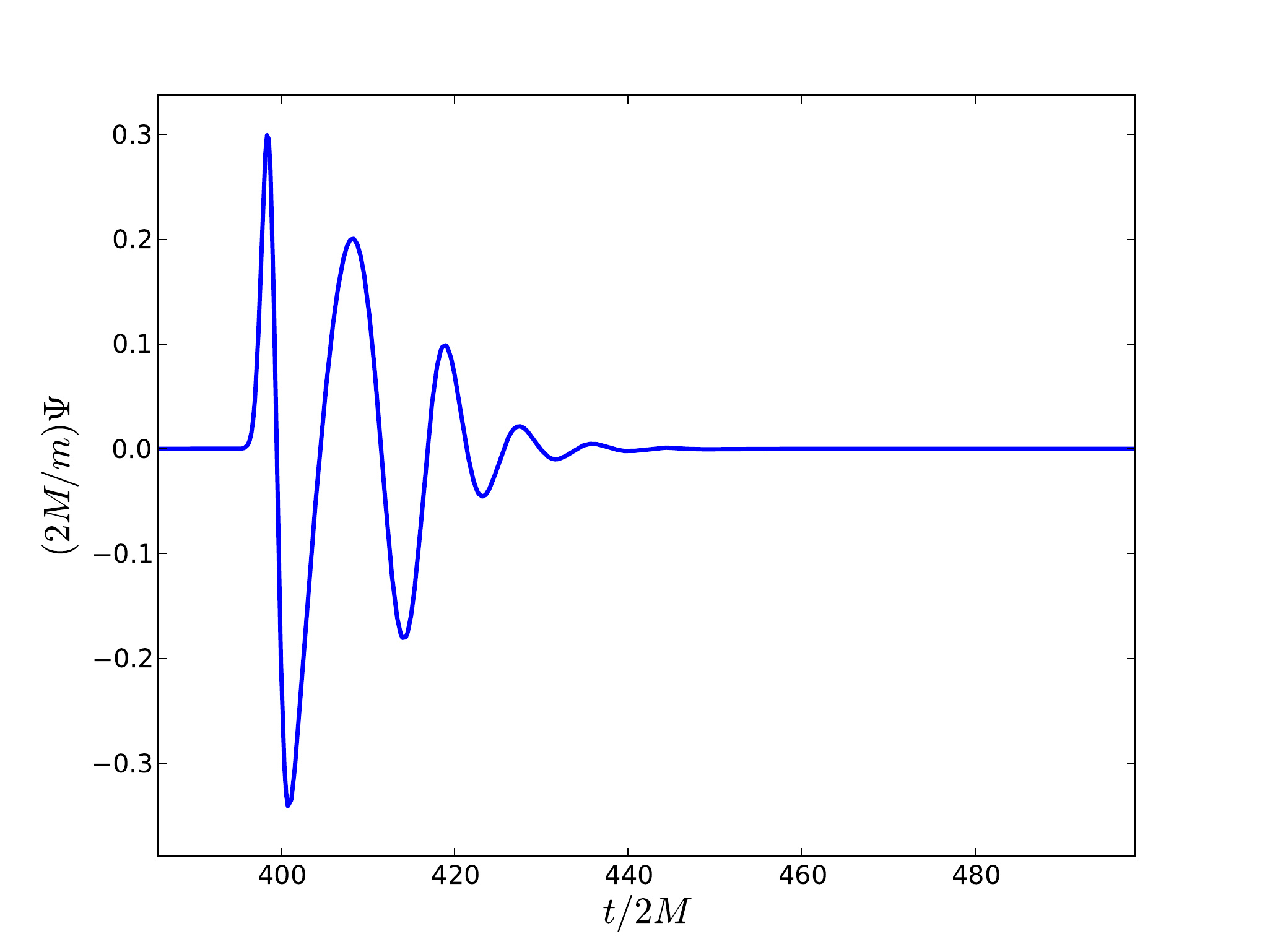}
\end{center}
\caption{\small The waveform, $r_{u0}=5(2M)$, of a Gaussian, time-symmetric initial pulse. The observer is located at $r^*=400(2M)$.}
\label{fig:Psi_h01}
\end{figure}

\begin{figure}[H]
\begin{center}
\includegraphics[width=8.5cm]{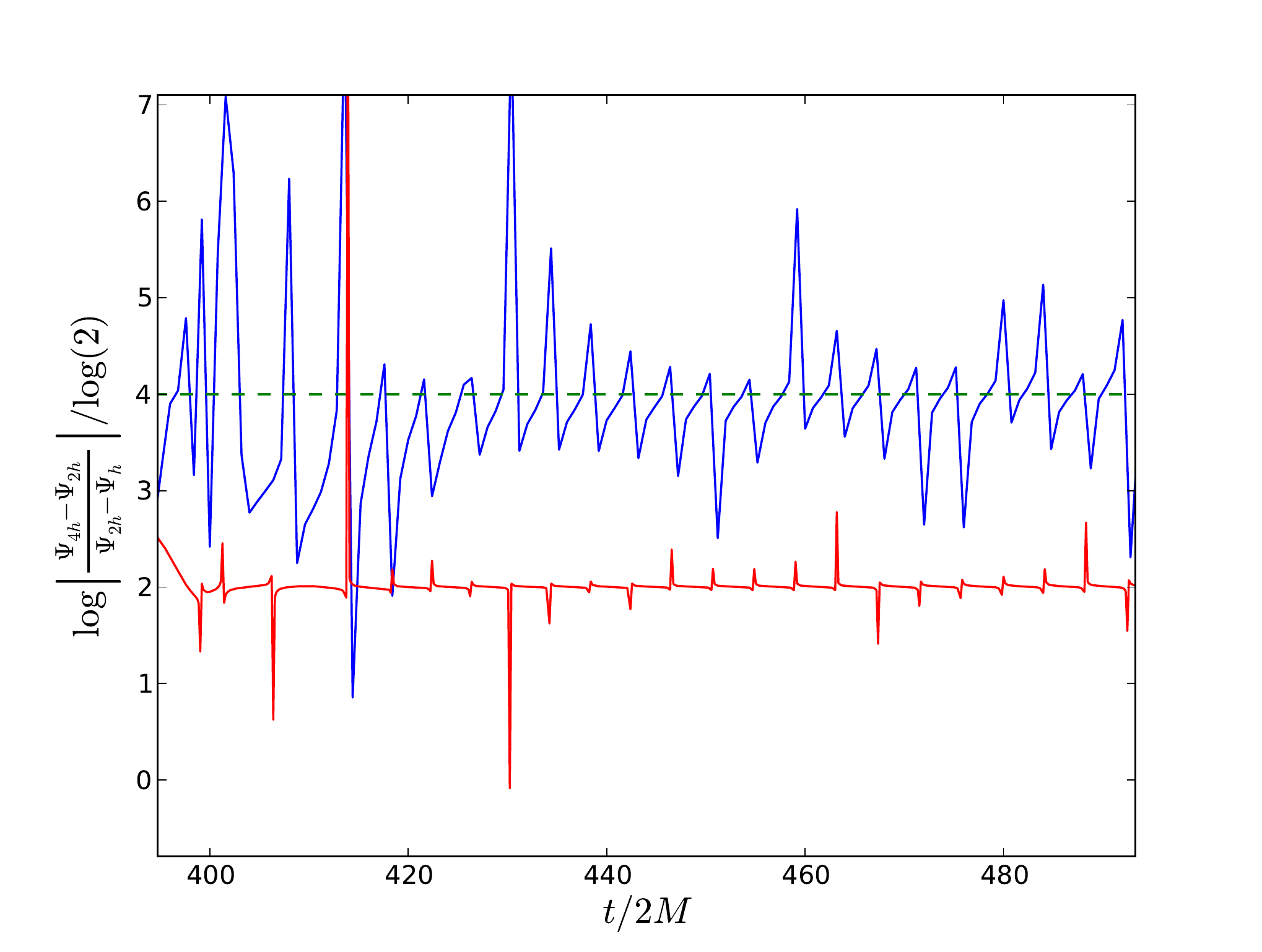}
\end{center}
\caption{\small Convergence rates of the fourth and second order algorithms, $r_{u0}=5(2M)$.}
\label{fig:conv_4th}
\end{figure}

Fig. \ref{fig:conv_4th}, obtained for $r_{u0}=5(2M)$, shows the fourth  and second order convergence rates (we remind that the first order code \cite{aosp11b} includes empty cells dealt at second order).

}

\section{Conclusions}

We have presented a fourth order novel integration method in time domain for the Zerilli wave equation. We have focused our attention to the even perturbations produced by a particle plunging in a non-rotating black hole.  
For cells crossed by the particle world line, the forward time wave function value at the upper node of the $(t, r^*$) grid cell is obtained by the combination of the preceding node values of the same cell, analytic expressions related to the jump conditions, { and} the values of the wave function at adjacent cells. In this manner, the numerical integration does not deal directly nor with the source term and the associated singularities, nor with the potential term. In short, the direct integration of the wave equation is avoided. { For other cells}, we refer instead to already published approaches {\cite{ha07}}.

The scheme has also been applied to circular and eccentric orbits and it will be object of a forthcoming publication.  

\section*{Acknowledgements}

{ The referees are thanked for careful reading and suggestions.} 
The authors wish to acknowledge the FNAK (Fondation Nationale Alfred Kastler), the CJC (Conf\'ed\'eration des Jeunes Chercheurs) and all organisations which stand against discrimination of foreign researchers.

{
\section*{Appendix}

Through Eq. \ref{scheme1_with_der}, we have determined the value of $\Psi$ at the upper node of the cell as function of the analytic jump conditions and of the time derivatives of the wave function up to fourth order. The derivatives are evaluated at the point $b$ and weighted by five coefficients $c_0$, $c_1$, $c_2$, $c_3$ and $c_4$. Afterwards, the derivatives are converted into a linear combination of the wave function values taken on points at the left and right sides of the trajectory. Indeed, Eq. \ref{syst_explicit} represents such a system of linear equations.
By injection of Eqs. \ref{devel.3}-\ref{devel.6} { into} Eq. \ref{syst_explicit}, we get

\beq
\fl\!\!\!\!\!\!\!\!
\begin{array}{c}
\mathcal{A}_2T_{\alpha_2}^{(0,0)}\Psi^-_b+\mathcal{A}_2T_{\alpha_2}^{(0,1)}\partial_t\Psi^-_b+\mathcal{A}_2T_{\alpha_2}^{(0,2)}\partial^2_t\Psi^-_b+\cdots+\mathcal{A}_2T_{\alpha_2}^{(1,3)}\partial_{r*}\partial^3_t\Psi^-_b\\
+\\
\mathcal{A}_4T_{\alpha_4}^{(0,0)}\Psi^-_b+\mathcal{A}_4T_{\alpha_4}^{(0,1)}\partial_t\Psi^-_b+\mathcal{A}_4T_{\alpha_4}^{(0,2)}\partial^2_t\Psi^-_b+\cdots+\mathcal{A}_4T_{\alpha_4}^{(1,3)}\partial_{r*}\partial^3_t\Psi^-_b\\
+\\
\vdots\\
+\\
\mathcal{N}^R_4T_{\nu^R_4}^{(0,0)}\Psi^-_b+\mathcal{N}^R_4T_{\nu^R_4}^{(0,1)}\partial_t\Psi^-_b+\mathcal{N}^R_4T_{\nu^R_4}^{(0,2)}\partial^2_t\Psi^-_b+\cdots+\mathcal{N}^R_4T_{\nu^R_4}^{(1,3)}\partial_{r*}\partial^3_t\Psi^-_b
\end{array}
=
\begin{array}{c}
c_0\Psi^-_b\\
+\\
c_1\partial_t\Psi^-_b\\
+\\
c_2\partial^2_t\Psi^-_b\\
+\\
c_3\partial^3_t\Psi^-_b\\
+\\
c_4\partial^4_t\Psi^-_b
\end{array}~,
\label{Ap_eq1a}
\eeq

\nid where $T_{p}^{(n,m)}$ represent the Taylor series coefficients at $p$ in the neighbourhood of $b$ and the indexes correspond to $n^{th}$ space and $m^{th}$ time derivatives. The wave function at $p$ is thus given by

\beq
\Psi^\pm_{p}=\sum_{n+m\leq4}T_{p}^{(n,m)}\partial_{r^*}^{n}\partial_{t}^{m}\Psi^\pm_b+{\cal O}\left(h^5\right)~.
\eeq

\nid An example shows the procedure which is applicable to all cases. We pick the node $\alpha_2$, Eq. \ref{devel.2}, where $T_{\alpha_2}^{(n,m)}=(-1)^n\frac{(2h-\epsilon_b)^n}{n!}$ and remind that  $T_{p}^{(0,0)}=1\ \forall\ p$. By grouping the derivatives, we get

\beq
\fl\qquad  
\begin{array}{c}
\left(\mathcal{A}_2T_{\alpha_2}^{(0,0)}+\mathcal{A}_4T_{\alpha_4}^{(0,0)}+\cdots+\mathcal{N}^R_4T_{\nu^R_4}^{(0,0)}\right)\Psi^-_b\\
+\\
\left(\mathcal{A}_2T_{\alpha_2}^{(0,1)}+\mathcal{A}_4T_{\alpha_4}^{(0,1)}+\cdots+\mathcal{N}^R_4T_{\nu^R_4}^{(0,1)}\right)\partial_t\Psi^-_b\\
+\\
\left(\mathcal{A}_2T_{\alpha_2}^{(0,2)}+\mathcal{A}_4T_{\alpha_4}^{(0,2)}+\cdots+\mathcal{N}^R_4T_{\nu^R_4}^{(0,2)}\right)\partial^2_t\Psi^-_b\\
+\\
\vdots\\
+\\
\left(\mathcal{A}_2T_{\alpha_2}^{(1,3)}+\mathcal{A}_4T_{\alpha_4}^{(1,3)}+\cdots+\mathcal{N}^R_4T_{\nu^R_4}^{(1,3)}\right)\partial_{r*}\partial^3_t\Psi^-_b\\
\end{array}
=
\begin{array}{c}
c_0\Psi^-_b\\
+\\
c_1\partial_t\Psi^-_b\\
+\\
c_2\partial^2_t\Psi^-_b\\
+\\
c_3\partial^3_t\Psi^-_b\\
+\\
c_4\partial^4_t\Psi^-_b
\end{array}~.
\label{Ap_eq1b}
\eeq

\nid By identification, we obtain a linear system, that is cast in the form

\beq
\fl
\underbrace{\left(
\begin{array}{ccccccccc}
1                    & \cdots & 1                    & \cdots & 1 & \cdots & 1 & \cdots & 1\\
T_{\alpha_1}^{(0,1)} & \cdots & T_{\beta^L_1}^{(0,1)}  & \cdots & T_{\gamma^L_2}^{(0,1)}  & \cdots & T_{\mu^L_3}^{(0,1)}  & \cdots & T_{\nu^R_4}^{(0,1)}\\
\vdots               & \vdots & \vdots               & \vdots & \vdots & \vdots & \vdots & \vdots\\
T_{\alpha_1}^{(0,4)} & \cdots & T_{\beta^L_1}^{(0,4)}  & \cdots & T_{\gamma^L_2}^{(0,4)}  & \cdots & T_{\mu^L_3}^{(0,4)} & \cdots & T_{\nu^R_4}^{(0,4)}\\
\vdots               & \vdots & \vdots               & \vdots & \vdots & \vdots & \vdots & \vdots\\
T_{\alpha_1}^{(1,0)} & \cdots & T_{\beta^L_1}^{(1,0)}  & \cdots & T_{\gamma^L_2}^{(1,0)}  & \cdots & T_{\mu^L_3}^{(1,0)}  & \cdots & T_{\nu^R_4}^{(1,0)}\\
\vdots               & \vdots & \vdots               & \vdots & \vdots & \vdots & \vdots & \vdots\\
T_{\alpha_1}^{(1,3)} & \cdots & T_{\beta^L_1}^{(1,3)}  & \cdots & T_{\gamma^L_2}^{(1,3)}  & \cdots & T_{\mu^L_3}^{(1,3)}  & \cdots & T_{\nu^R_4}^{(1,3)}
\end{array}
\right)}_{\mathbb{T}}
\underbrace{\left(
\begin{array}{c}
\mathcal{A}_2\\
\vdots\\
\mathcal{B}^L_1\\
\vdots\\
\mathcal{G}^L_2\\
\vdots\\
\mathcal{M}^L_3\\
\vdots\\
\mathcal{N}^R_4
\end{array}
\right)}_{\mathbb{P}}
=
\underbrace{\left(
\begin{array}{c}
c_0\\
c_1\\
c_2\\
c_3\\
c_4\\
0\\
\vdots\\
0\\
\vdots\\
0
\end{array}
\right)}_{\mathbb{C}}~,
\eeq

\nid where the upper indexes $(n,m)$ cover all combinations such that $n\!+\!m\leq4$. Finally, by inversion of the $\mathbb{T}$ matrix, the unknown terms of the $\mathbb{P}$ vector are identified.
}

\end{document}